\documentclass{aa}  

\usepackage{graphicx}
\usepackage{txfonts}
\usepackage{amsmath}	
\usepackage{amssymb}	
\usepackage[dvipsnames]{xcolor}
\usepackage{rotating}
\usepackage{url}
\usepackage{caption}
\usepackage[colorlinks=true,citecolor=blue,linkcolor=blue]{hyperref}
\newcommand{\chandra}{\textit{Chandra}}

\newcommand{\athena}{{\it Athena}}
\newcommand{\xrism}{{\it XRISM}}

\newcommand{\src}{Cyg~X-1}

\newcommand{\fvsq}{$F_\mathrm{var}^2$}
\newcommand{\fvsqbin}{$F_\mathrm{var,\ binning}^2$}

\begin{document}

   \title{Stellar wind variability in Cygnus~X-1 from high-resolution excess variance spectroscopy with Chandra}
   \titlerunning{Stellar wind excess variance in Cyg X-1 with Chandra}
   \authorrunning{L. H\"arer et al.}

   \author{L. K. H\"arer\inst{1,2}\fnmsep\thanks{\email{lucia.haerer@fau.de}}, 
            M. L. Parker\inst{3},
             I. El Mellah\inst{4,5},
            V. Grinberg\inst{6},
            R. Ballhausen\inst{7,8},
            Z. Igo\inst{9},
            A. Joyce\inst{1},
            \and
            J. Wilms\inst{1}}

   \institute{Dr.~Karl Remeis Sternwarte \& Erlangen Centre for Astroparticle Physics, Friedrich-Alexander-Universität Erlangen-Nürnberg,
Sternwartstr. 7, 96049 Bamberg, Germany
            \and
            Max-Planck-Institut für Kernphysik, Saupfercheckweg 1, 69117 Heidelberg, Germany
            \and
            Institute of Astronomy, Madingley Road, Cambridge, CB3 0HA, UK
            \and
            Institut de Plan\'{e}tologie et d'Astrophysique de Grenoble, 414 Rue de la Piscine, 38400 Saint-Martin-d'Hères, France
            \and
            Departamento de Física, Universidad de Santiago de Chile, Av. Victor Jara 3659, Santiago, Chile
            \and
             European Space Agency (ESA), European Space Research and Technology Centre (ESTEC), Keplerlaan 1, 2201 AZ Noordwijk, The Netherlands
            \and
            Department of Astronomy, University of Maryland, College Park, MD 20742
            \and
            Center for Research and Exploration in Space Science and Technology, NASA/GSFC, Greenbelt, MD 20771
            \and
            Max-Planck-Institut für Extraterrestrische Physik (MPE), Giessenbachstrasse 1, 85748 Garching bei München, Germany
             }

   \date{Received 17/04/2023; accepted 19/09/2023}

 
  \abstract
   {Stellar winds of massive stars are known to be driven by line absorption of UV photons, a mechanism which is prone to instabilities, causing the wind to be clumpy. The clumpy structure hampers wind mass-loss estimates, limiting our understanding of massive star evolution. The wind structure also impacts accretion in high-mass X-ray binary (HMXB) systems.}
   {We analyse the wavelength-dependent variability of X-ray absorption in the wind to study its structure. Such an approach is possible in HMXBs, where the compact object serves as an X-ray backlight. We probe different parts of the wind by analysing data taken at superior and inferior conjunction.}
   {We apply excess variance spectroscopy to study the wavelength-dependent soft ($2\mbox{--}14$\AA) X-ray variability of the HMXB Cygnus~X-1 in the low/hard spectral state. Excess variance spectroscopy quantifies the variability of an object above the statistical noise as a function of wavelength, which allows us to study the variability of individual spectral lines. As one of the first studies, we apply this technique to high-resolution gratings spectra provided by \chandra, accounting for various systematic effects. The frequency dependence is investigated by changing the time binning. }
   {The strong orbital phase dependence we observe in the excess variance is consistent with column density variations predicted by a simple model for a clumpy wind. We identify spikes of increased variability with spectral features found by previous spectroscopic analyses of the same data set, most notably from silicon in over-dense clumps in the wind. In the silicon line region, the variability power is redistributed towards lower frequencies, hinting at increased line variability in large clumps. In prospect of the microcalorimetry missions that are scheduled to launch within the next decade, excess variance spectra present a promising approach to constrain the wind structure, especially if accompanied by models that consider changing ionisation.}
   {}
   

   \maketitle
%

\section{Introduction}
\label{sec:intro}

Accretion onto compact objects is a dynamically complex phenomenon and releases enormous amounts of highly-energetic radiation that strongly impacts the environment of the accretor through feedback processes. The complex dynamics and, in particular, inhomogeneities in the accretion flow often induce variability, which can span many orders of magnitude \citep[e.g.,][]{Oskinova12}. Understanding accretion processes on all scales, in galactic and extra-galactic sources, therefore requires precise measurements of both the variability and the environmental characteristics (e.g., densities, ionisations, and element abundances), which calls for a joined spectral and timing analysis. 

Excess variance spectroscopy excels at combining spectral and timing information. The excess variance is the variance of a light curve above the expected statistical noise \citep{Edelson02, Vaughan03}. If applied to a set of light curves taken in different bands, the spectral distribution of the variability can be studied. Excess variance spectroscopy can reveal the variability behaviour of individual spectral lines, provided sufficient resolution and signal. This presents a clear advantage over other tools, such as colour-colour diagrams \citep{Nowak_2011a,Hirsch19,Grinberg_2020a}. So far, excess variance spectra have mainly been applied to study extra-galactic sources \citep[see, e.g., references in][]{Parker20}. In particular, \citet{Parker17_iras, Parker18_PCA} introduced a method to detect ultra-fast outflows of active galactic nuclei (AGN), which are fast wide-angle accretion disk winds. They possess a large amount of mechanical power and therefore influence AGN evolution via feedback processes \citep[e.g.,][]{Tombesi10}. In recent years, excess variance spectroscopy has proven valuable in detecting ultra-fast outflows and in independently confirming and refining known results \citep{Parker17_iras, Parker18_PCA, Igo20, Haerer21_PDS}, demonstrating their advantages: excess variance spectra are easy to calculate, provide less biased detections, and show more pronounced features than count spectra. Nevertheless, they have seldom been applied to galactic sources, where high-resolution gratings data present an opportunity to study the wavelength dependence of variability at a resolution above $0.1\AA$. We demonstrate the capabilities of excess variance spectroscopy for studying the structure of stellar winds by applying it to \chandra\ observations of the High-Mass X-ray Binary (HMXB) Cygnus~X-1 (\src). 

\begin{figure*}
    \centering
    \includegraphics[width=\textwidth]{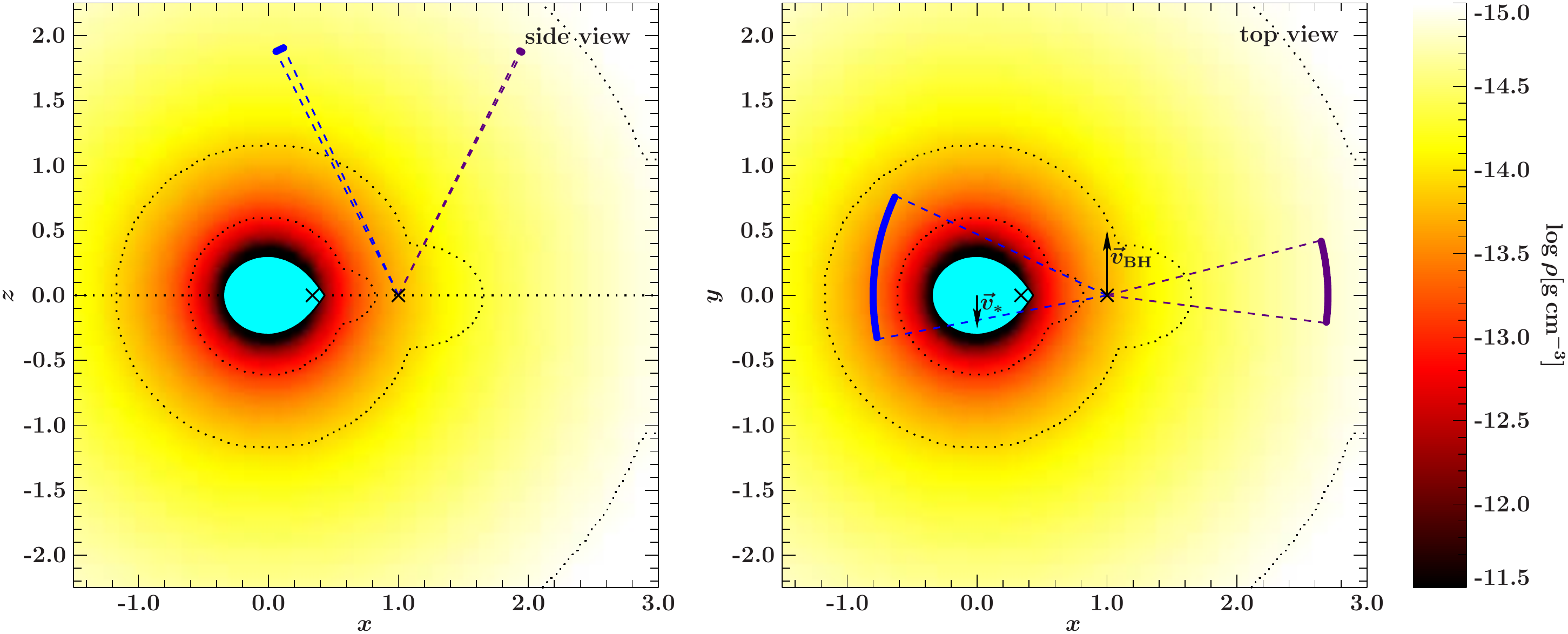}
    \caption{The density of the stellar wind in \src\ according to the isotropic clumpy wind model by \citet{ElMellah20}. The average density is shown for clarity, instead of the clumpy wind structure. The dotted contours denote densities of $\log \rho/(\mathrm{g}\, \mathrm{cm}^{-3}) = -13$, -14, and -15 in the focused wind model by \citet{Gies86b}. The left panel shows a side view (dotted line: orbital plane) and the right a top view (arrows: orbital motion). The line of sight to the black hole and orbital phase covered is highlighted for Obs.~3814 (dashed, blue) and 11044 (dashed, purple). Masses and orbital separation were set according to \citet{Miller-Jones21}.}
    \label{fig:cyg_density}
\end{figure*}

As binary systems consisting of an O or B star and neutron star or black hole, HMXBs are invaluable laboratories to study accretion, compact objects, and the strong winds of massive stars. \src\ is one of the most studied HMXBs. As a bright and persistent source, it was discovered in 1964 \citep{Bowyer65}. The compact object is dynamically constrained to be a black hole \citep{Gies82} and recently, \citet{Miller-Jones21} refined the mass estimate to $21.2\pm2.2\,\mathrm{M}_\odot$, making \src\ the most massive known stellar mass black hole in an X-ray binary. The companion star \object{HDE~226868} is an O9.7 Iab supergiant \citep{Walborn73} with a mass of ${\sim}41\,\mathrm{M}_\odot$ \citep{Miller-Jones21}. As a blue supergiant star, HDE~226868 has a strong stellar wind with a mass-loss rate of ${\sim}10^{-6}\, \mathrm{M}_\odot\,\mathrm{yr}^{-1}$ \citep{Herrero95, Gies03} and a terminal velocity of ${\sim}2100\,\mathrm{km}\,\mathrm{s}^{-1}$ \citep{Herrero95}. Figure~\ref{fig:cyg_density} depicts the system and wind geometry of \src. The system is seen under an inclination of $27\mbox{--}28^\circ$\citep{Orosz11}. Due to the low orbital separation \citep[0.244\,AU,][]{Miller-Jones21} and short orbital period \citep[$5.6\,$d,][]{Webster72, Brocksopp99, Gies03}, the donor star is close to filling its Roche lobe, which gives rise to a focused accretion stream along the line connecting the supergiant and the black hole. This geometry was first verified in the optical by \citet{Gies86a, Gies86b}, with an analysis based on \citet{Friend82}, and is consistent with X-ray measurements of, e.g., the orbital phase dependence of the column density \citep[see][and references therein]{Miskovicova16}. 

Winds of massive stars, such as HDE~226868 in \src, are driven by resonant line-absorption of stellar UV photons by partly ionised metal ions \citep{Lucy70, Castor75}. This physical mechanism is prone to the line-deshadowing instability whose development produces over-dense small scale regions called clumps \citep{Owocki84, Owocki88, Sundqvist18a}. Most of the wind mass is contained in the clumps, making these winds notoriously inhomogeneous \citep{Hamann08, Sundqvist12, Puls15}. Our limited knowledge of wind clumpiness hampers the accuracy of the mass loss rates deduced from all available diagnostics \citep{Fullerton06, Sundqvist18b}. Mass loss, however, plays a key role in the evolution of massive stars, especially in the late stages \citep{Puls08}. Consequentially, the inaccuracy of mass loss diagnostics limits our understanding of the role radiative, mechanical, and chemical feedback of massive stars plays in many contexts.
In HMXBs specifically, the mass loss and the wind structure influence the accretion flow. 
\citet{in'tZand05} first suggested that, in HMXBs, clumpiness could be constrained from the variability of the mass accretion rate onto the compact objects. This claim was challenged by \citet{ElMellah18} who showed that accreted clumps tend to mix inside the shocked region surrounding the compact object, which smears out the variability induced by stochastic clump capture. Alternatively, the variability of the column density induced by unaccreted wind clumps passing by the line-of-sight can be used \citep{Oskinova12}. Motivated by \citet{Grinberg15}, who constrained clump properties in \src\ with this ansatz, \citet{ElMellah20}, EG20 hereafter, devised a model that connects the statistical properties of the column density variations to the clump size and mass, i.e., the model quantitatively links variability and clump properties.

We apply excess variance spectroscopy to study the stellar wind variability in \src. Our aim is both, to test the results obtained previously with an independent approach and to establish excess variance spectroscopy both as an approach for wind studies in HMXBs and as a tool that can be used for high resolution X-ray data, such as obtainable today with the High-Energy Transmission Grating (HETG) on-board \chandra , or will become available with the launch of the \textsl{XRISM} and \textsl{Athena} X-ray missions. We analyse \chandra/HETG observations at superior and inferior conjunction of which detailed spectral analysis has been performed in \citet{Hanke09}, \citet{Miskovicova16}, and \citet{Hirsch19}. The existing studies constitute an excellent test case for constraining the stellar wind structure with excess variance spectroscopy and are summarised in Sect.~\ref{sec:cyg_clumps}. We take advantage of the predictions of column density variability expected in such a system recently introduced by EG20. Section~\ref{sec:obs_meth} provides more details on the observations, the calculation of the excess variance spectra, the treatment of systematic variability, and the selection of lines we check against. The resulting excess variance spectra and the frequency dependence of the excess variance are shown in Sect.~\ref{sec:resluts}. In Sect.~\ref{sec:discussion}, we explore the driving mechanism behind the observed variability, model the strong orbital phase dependence based on EG20, and discuss implications for the wind geometry and structure and the role of varying ionisation, before summarising our conclusions in Sect.~\ref{sec:concl}.

\section{The clumpy wind in \src}
\label{sec:cyg_clumps}

Highly variable absorption is observed from \object{Cyg X-1} especially in the low/hard state of the black hole \citep[e.g.,][]{Feng_2002a,Ibragimov_2005a,Boroson_2010a,Grinberg15}\footnote{In the low/hard state, the intrinsic emission from close to the black hole is dominated by a hard, Comptonised component, with comparatively little contribution from the thermal emission of the accretion disk \citep[for a discussion of accretion states in Cyg~X-1 see, e.g.,][]{Wilms_2006a,Grinberg13}.}. In particular, dips in the light curve are observed in soft X-rays preferentially at superior conjunction and interpreted as signatures of the clumps in the wind \citep[see][and references therein]{Grinberg15}. At superior conjunction, the line of sight passes closer to the star, potentially sampling a highly structured wind region (see Fig.~\ref{fig:cyg_density}). 

A paper series by \citet{Hanke09}, \citet{Miskovicova16}, and \citet{Hirsch19} used high-resolution \chandra\ spectra to study the wind in the low/hard state in great detail. \citet{Hanke09} investigated the non-dip spectrum at superior conjunction, identifying absorption lines from several H- and He-like ions and L-shell iron at low velocity shifts. \citet{Miskovicova16} expanded this study to other orbital phases. They observed P-Cygni line profiles at inferior conjunction, which point to weak absorption and a high projected velocity. As the last paper in the series, \citet{Hirsch19} investigated the above mentioned absorption dips in the light curve, which are believed to be caused by clumps in the wind. Using time resolved spectra, they studied how absorption from silicon and sulphur ions evolves in the dips, finding that lower ionisation species appear and increase in relative strength with dip depth. From this result, \citet{Hirsch19} concluded that material in the clumps has a lower ionisation than its surroundings and that deeper dips correspond to larger clumps. In other words, the clumps have a layered ionisation structure, with large clumps reaching lower ionisations in their centres. An open question that remains from these studies is whether the dips are caused by individual, big clumps passing through the line of sight or by groups of several smaller ones. The geometry (spherical or "pancake"-like) and typical masses also remain to be constrained. 

\section{Observations and Methods}
\label{sec:obs_meth}

\subsection{Chandra HETG observations of \src}
\label{sec:obs}

We analyse two \chandra\ observations (ObsIDs 3814 and 11044) of \src\ in the low/hard state taken with the High-Energy Transmission Grating \citep[HETG,][]{Canizares05} and the Advanced CCD Imaging Spectrometer \citep[ACIS,][]{Garmire03}. The observations cover the superior (3814, orbital phase $\varphi=0.93\mbox{--}0.03$) and inferior (11044, $\varphi=0.48\mbox{--}0.54$) conjunction passages, during which the line of sight probes different parts of the wind (see Fig.~\ref{fig:cyg_density}). At superior conjunction, the line of sight intercepts the wind close to the stellar photosphere and grazes the focused stream. At inferior conjunction, the wind is probed at a greater distance to the companion. Note that the low/hard classification of these observations follows \citet{Grinberg13} and the orbital phases were assigned according to the ephemeris provided by \cite{Gies03}. 

Detailed spectral analysis of both data sets has been performed in the studies mentioned in Sect.~\ref{sec:cyg_clumps} \citep{Hanke09, Miskovicova16, Hirsch19}. The existing results provide an opportunity to assess the capabilities of our approach against those of conventional analysis techniques. The observations were performed in timed exposure mode (TE), i.e., events are accumulated on the CCDs, transferred in the frame store, and read out. A full read-out cycle of the ACIS-S in TE mode usually takes 3.2$\,$s, however, in our case the time is halved to 1.7$\,$s, because only 512 of the 1024 available pixel rows were in use. We use the extracted event files provided by \citet{Miskovicova16}, but analyse the entirety of Obs.~3814, including dipping and non-dipping stages. The extractions were created using CIAO version 4.2, consistently with previous work by \citet{Miskovicova16} and \citet{Hirsch19} in order to facilitate direct comparison with their spectral results. We include all four first order spectra from the High and Medium Energy Gratings (HEG and MEG, orders $\pm1$) of the HETG. Refer to \citet{Miskovicova16} and \citet{Hirsch19} for further details on the data taking, extraction, and processing.

We perform a spectral and time binning on the extracted events, separately for the four first orders of the HEG and MEG. The spectral binning covers the $2\mbox{--}14$\AA\ band with a 0.05\AA\ resolution and a $500\,$s step is used for the time binning. For a closer inspection later on, we increase the spectral resolution threefold and vary the time step in the $50\mbox{--}2000\,$s range. The excess variance spectrum is then calculated as described in Sect.~\ref{sec:fvsq}. While the spectral resolution and smallest time step are constrained by the signal strength, the maximal time step should be chosen such that the light curve has a sufficient number of bins to obtain an accurate estimate of the variance \citep[$\gtrsim 20$,][]{Vaughan03}. One sigma uncertainties are given unless otherwise noted.

\subsection{Calculating the excess variance spectrum}
\label{sec:fvsq}

The excess variance quantifies the variability of an object and is defined as \citep{Edelson02, Vaughan03}:
\begin{equation}
    \label{equ:exvar}
    \sigma^2_\mathrm{XS} = S^2 - \overline{\sigma_\mathrm{err}^2} \, , 
\end{equation}
where $S^2$ is the variance of the light curve and $\overline{\sigma_\mathrm{err}^2}$ the arithmetic mean of the squared measurement uncertainty, i.e., the variability introduced by the statistical nature of the measurement process. Subtracting this statistical noise makes the excess variance a quantity which is indicative of only the physical variability of the source, in other words, if no variability is present, the excess variance is zero. The excess variance depends on source flux, which is avoided by normalising it to the square of the mean number of counts per bin, $\bar{x}$:
\begin{equation}
    \label{equ:fvar}
    F_\mathrm{var}^2 = \frac{\sigma^2_\mathrm{XS}}{ \bar{x}^2 } \, .
\end{equation}

We write the normalised excess variance in terms of the square of the fractional quantity, $F_\mathrm{var}$, also called root mean square (RMS) variability. $F_\mathrm{var}$ is often used as it is linear in number of counts. However, the squared quantity, \fvsq, is preferable if the variability is low, because $F_\mathrm{var}$ contains a square root and becomes imaginary if the uncertainty, $\overline{\sigma_\mathrm{err}^2}$, is larger than the signal, $S^2$. This case is relevant to our analysis because working at gratings resolution makes it necessary to divide the signal over a large number of bins, in comparison to CCD data, to which the excess variance is usually applied.

One advantage of using the excess variance as a measure for variability is the straightforward application to spectral data, i.e., the calculation of an excess variance spectrum. As discussed in the introduction, studying the wavelength dependence of variability can provide unique physical insights that cannot be obtained from static count spectra. To judge the significance of features in the excess variance spectrum, an error estimate is required. We use the expression computed by \citet{Vaughan03} from a Monte Carlo approach, which accounts for Gaussian and Poissonian measurement uncertainties,
\begin{equation}
    \mathrm{err}(F_\mathrm{var}^2) = \sqrt{\frac{2\overline{\sigma^2_\mathrm{err}}}{N\bar{x}^2}\left( \frac{\overline{\sigma^2_\mathrm{err}}}{\bar{x}^2}+ 2F_\mathrm{var}^2 \right)} \, ,
\end{equation}
where N is the number of photons in the respective bin.

The band pass of frequencies contributing to the excess variance is determined by the time binning of the light curve and the length of the observation. By choosing the time binning, it is therefore possible to probe the frequency dependence of \fvsq\ and investigate if there is a variability timescale intrinsic to the system. \citet{Hirsch19} state that the passage of a clump across the line of sight takes $0.5\mbox{--}5\,$ks, assuming that the observed dips in the spectrum are caused by single clumps and not multiple small clumps. We choose a time step of 500$\,$s for our initial investigation to capture all clump variability in this timescale range. The frequency dependence of the detected features is then investigated in a second step.

\subsection{Eliminating systematic variability}
\label{sec:sysvar}

As detailed in Sect.~\ref{sec:obs}, we analyse \chandra\ HETG data taken with the ACIS-S detector. For a variability study such as that presented in this work, a solid understanding of the timing and read-out properties of the detector and associated systematic effects is vital. 

The ACIS-S detector consists of a row of six CCDs, which are read out in sequence. As a consequence, photons arriving simultaneously on different CCDs are registered at slightly different times. This effect becomes relevant if data from multiple gratings or orders, i.e., locations on the detector, are combined. We avoid systematic effects of this kind by calculating \fvsq\ separately for each grating and order, $i$, and taking their weighted mean,  
\begin{equation}
    F_\mathrm{var}^2 = \frac{1}{\sum_i \bar{x}^2_i} \sum_i F_{\mathrm{var}, \, i}^2\cdot\bar{x}^2_i \, .
\end{equation}

Weighting by the square of the mean number of counts, $\bar{x}^2_i$, gives equal weight to each count, regardless of which grating or order it belongs to. As the errors $\mathrm{err(}F_{\mathrm{var}, \, i}^2)$ are independent, they can be propagated according to 

\begin{equation}
    \mathrm{err}({F_\mathrm{var}^2}) = \frac{1}{\sum_i \bar{x}^2_i} \sqrt{\sum_i \bar{x}^4_i \cdot \mathrm{err}({F_{\mathrm{var},\, i}^2})^2} \, .
\end{equation}

A second systematic effect is the increase in \fvsq\ due to the photon loss in gaps between CCDs and rows of dead pixels in \chandra's ACIS-S detector. The photon loss is averaged over a range of wavelengths by dithering the telescope pointing\footnote{ \url{https://cxc.cfa.harvard.edu/proposer/POG/html/ACIS.html} (The \chandra\ Proposer's Observatory Guide, Sect.~6.12)}, which causes the count rate to vary periodically within dithering range of CCD gaps and dead rows, increasing the variability in this range. In the \fvsq\ spectrum, this increase manifests itself as spikes at specific wavelengths, which are not per se distinguishable from genuine features. Gap induced \fvsq\ spikes constitute the largest systematic effect in our analysis, exceeding genuine features up to about an order of magnitude. We detect and filter out affected wavebands separately for each grating and order, before calculating and averaging their $F_{\mathrm{var}, \, i}^2$ as described above. The resulting reductions in sensitivity strongly vary between wavelength bands. For a detailed description, see Appendix~\ref{sec:gapfiltering}.

While gaps produce spikes at certain wavelength, the discrete nature of the read-out increases the continuum variability for binned spectra. From Monte Carlo simulations detailed in Appendix~\ref{sec:binvar}, we conclude that the effect on our analysis is negligible but can be significant on timescales approaching the read-out time of the instrument. 

\subsection{Line selection}
\label{sec:fe-select}

We compiled atomic data from various sources to compare to observed features (see Sect.~\ref{sec:resluts}). Most of the data was taken from \texttt{ATOMDB}\footnote{\url{http://www.atomdb.org}} and \texttt{XSTARDB} \citep{Mendoza21}. The full list of references is shown in Table~\ref{tab:spikes_3814}. We only consider transitions from astrophysically abundant elements (e.g., C, N, O, Ne, Mg, Si, S, Fe etc.), which include the ground state. The final selection encompasses 719 lines with energies of $0.1\mbox{--}14.7\,$keV. For more than one transition at a given wavelength, line identification is based on elemental abundance, oscillator strength, and initial state population. However, fine structure splitting cannot be resolved for most of the lines.

\section{Results}
\label{sec:resluts}
\subsection{Overview: \fvsq\ continuum and spike features}
\label{sec:res_overview}

\begin{figure}
    \centering
    \includegraphics[width=\linewidth]{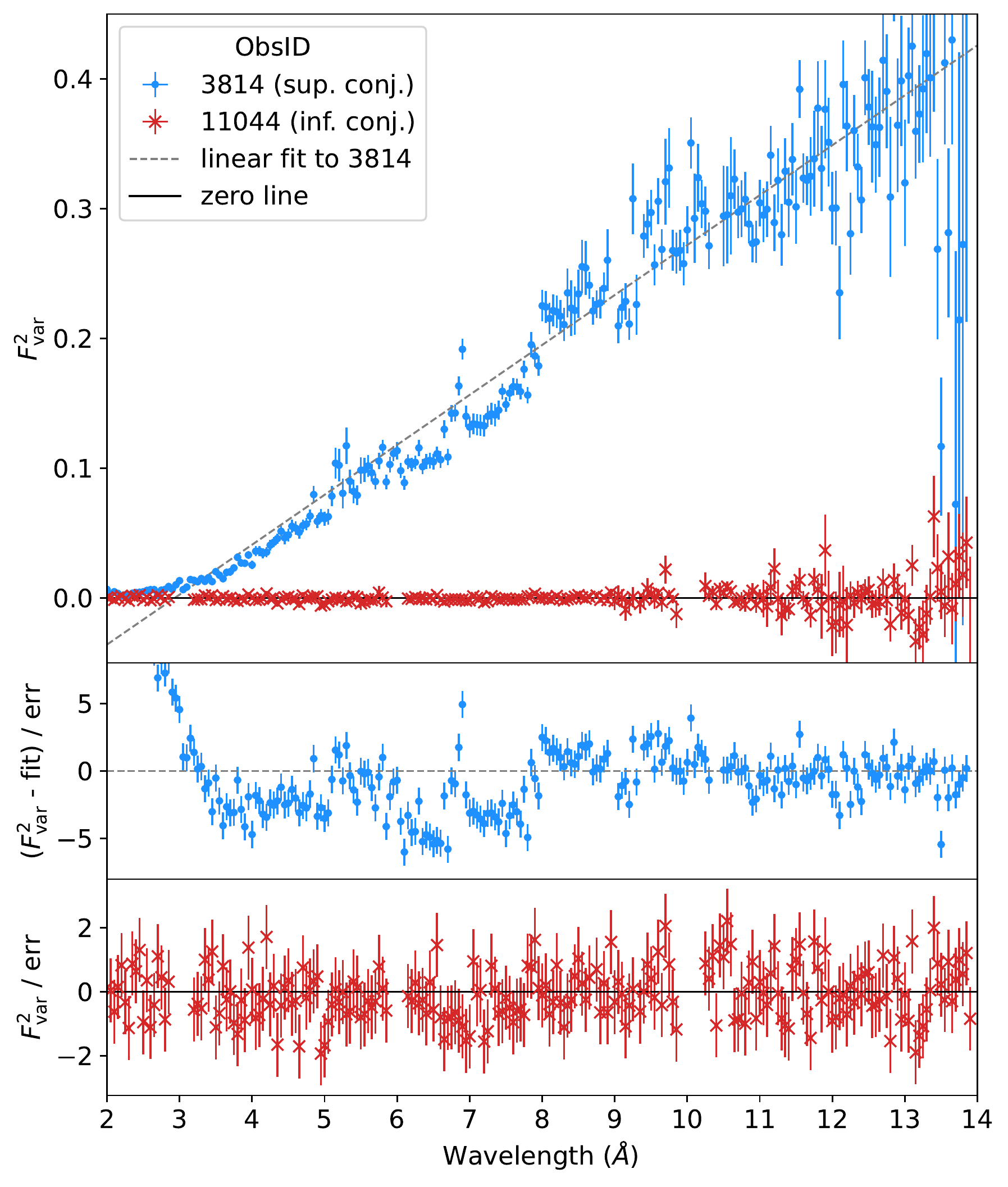}
    \caption{The \fvsq\ spectrum of Obs.~3814 (blue) and 11044 (red) during which \src\ passed through superior and inferior conjunction, respectively. Residuals to a continuum estimate (a linear fit for 3814, the zero-line for 11044) are shown in the bottom panels. The gaps in the \fvsq\ spectrum of Obs.~11044 around 3\AA, 6\AA, and 10\AA\ are due to the filtering of systematic variability from CCD gaps in \chandra's HETG detector (see Sect.~\ref{sec:sysvar} and Appendix~\ref{sec:gapfiltering}).}
    \label{fig:rms}
\end{figure}

\begin{sidewaysfigure*}
    \centering
    \includegraphics[width=0.9\textheight]{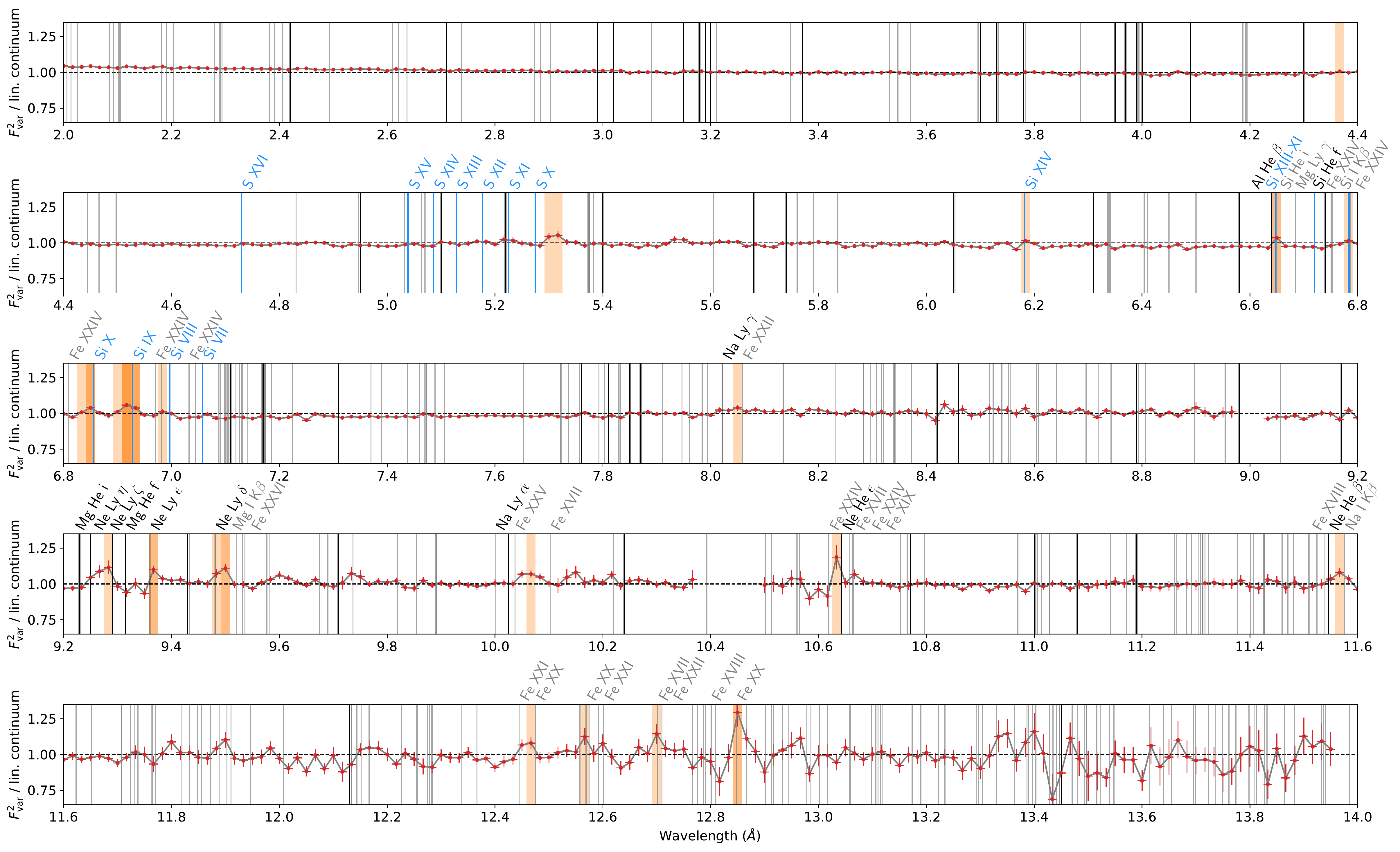}
    \caption{The \fvsq\ spectrum of Obs.~3814, relative to the continuum fit from Fig.~\ref{fig:rms}. Three sets of lines are shown: H and He-like absorption lines identified by \citet{Hanke09} (black), S and Si lines from \citet{Hirsch19} (blue), transitions from the selection described in Sect.~\ref{sec:fe-select} (grey), which includes iron. Spikes with a significance $\sigma > 2\sigma$ are highlighted in shades of orange according to the value of $\sigma$. Lines marked in black and grey are labelled only if they are close (${<}0.05$\AA) to a significant spike. For subsequent iron transitions belonging to the same ion the label is shown only once. A list of spikes and associated transitions can be found in Tab.~\ref{tab:spikes_3814}.}
    \label{fig:3814}
\end{sidewaysfigure*}

\defcitealias{Bearden67}{2}
\defcitealias{Drake88}{3}
\defcitealias{Garcia65}{4}
\defcitealias{Hell16}{5}
\defcitealias{Johnson85}{6}
\defcitealias{Liao13}{7}
\defcitealias{Verner96}{8}

\defcitealias{Hanke09}{a}
\defcitealias{Hirsch19}{b}

\begin{table*}
	\caption{List of spike features in the \fvsq\ spectrum of Obs.~3814 with a significance ${>}2\sigma$ and associated lines, which are defined as having a distance of ${<}0.05$\AA\ to the features. The listed lines are compiled the selection described in Sect.~\ref{sec:fe-select} and previous analyses of the data: \citepalias{Hanke09} \citealt{Hanke09} and \citepalias{Hirsch19} \citealt{Hirsch19}. Primary references are numbered and given below.}
	\begin{tabular}{c c l l c c}
	    \hline
		\hline
		Spike [\AA] & $\sigma$ & Ion & Line(s) & Wl. [\AA] & Ref. \\ 
		\hline
		4.37   & 2.0 & & & &\\
		\hline
		5.30   & 2.0 & \ion{S}{x} & L-shell & 5.275 & \citetalias{Hell16}, \citetalias{Hirsch19}\\
		5.32   & 2.3 & &&&\\
		\hline
		6.18   & 2.5 & \ion{Si}{xiv} & Ly $\alpha$ & 6.182 & \citetalias{Johnson85}, \citetalias{Hirsch19}\\ 
		\hline
		6.65   & 3.8 & \ion{Si}{xiii} & i [em.] & 6.685 & \citetalias{Drake88}, \citetalias{Hirsch19} \\ 
		&& \ion{Si}{xiii} & He $\alpha$ & 6.648 & \citetalias{Drake88}, \citetalias{Hirsch19}\\
		&& \ion{Al}{xii}  & He $\beta$ & 6.64 & \citetalias{Hanke09}\\ 
		\hline 
		6.78 & 2.8 & \ion{Fe}{xxiv} & 2p $\rightarrow$ 7d & 6.809  &  1\\
		&&\ion{Fe}{xxiv} & 2s $\rightarrow$ 6p & 6.787  &  1 \\
		&& \ion{Si}{xi} & L-shell & 6.784 & \citetalias{Hell16}, \citetalias{Hirsch19}\\
		&&\ion{Si}{i} & K$\beta$ & 6.753 & \citetalias{Bearden67} \\ 
		&&\ion{Fe}{xxiv} & 2p $\rightarrow$ 7d & 6.751 &  1\\
		&&\ion{Si}{xiii} & f [em.] & 6.740 & \citetalias{Drake88}, \citetalias{Hanke09}\\
		&&\ion{Mg}{xii} & Ly $\gamma$ & 6.738 & \citetalias{Garcia65}\\ 
		\hline
		6.83   & 2.3 & \ion{Si}{x} & L-shell & 6.856 & \citetalias{Hell16}, \citetalias{Hirsch19}\\
		6.85   & 4.5 & \ion{Fe}{xxiv} &2p $\rightarrow$ 7d & 6.809  &  1\\
		&& \ion{Fe}{xxiv} & 2s $\rightarrow$ 6p & 6.787  &  1\\
		\hline
		6.90   & 2.5 & \ion{Fe}{xxiv} & 2p $\rightarrow$ 6s & 6.982  &  1\\
		6.92   & 5.7 & \ion{Fe}{xxiv} & 2p $\rightarrow$ 6d & 6.970  &  1\\
		6.93   & 4.4 & \ion{Si}{ix} & L-shell & 6.928 & \citetalias{Hell16}, \citetalias{Hirsch19}\\
		\hline 
		6.98   & 2.1 & \ion{Fe}{xxiv} & 2p $\rightarrow$ 6d & 7.033  &  1\\
		&& \ion{Si}{viii} & L-shell & 6.996 & \citetalias{Hell16}, \citetalias{Hirsch19}\\
		&& \ion{Fe}{xxiv} & 2p $\rightarrow$ 6s & 6.982  &  1\\
		&& \ion{Fe}{xxiv} & 2p $\rightarrow$ 6d & 6.970 &  1\\
		\hline
		8.05   & 2.2 & \ion{Fe}{xxii} & 2p $\rightarrow$ 5d & 8.059  &  1\\
		&&\ion{Na}{xi} & Ly $\gamma$ & 8.021 & \citetalias{Garcia65}, \citetalias{Hanke09}\\ 
		\hline
		9.28   & 2.2 & \ion{Mg}{xi} & f [em.] & 9.314 & \citetalias{Drake88}, \citetalias{Hanke09}\\ 
		&& \ion{Ne}{x} & Ly $\zeta$ & 9.29 & \citetalias{Hanke09} \\
		&& \ion{Ne}{x} & Ly $\eta$ & 9.25 & \citetalias{Hanke09} \\
		&& \ion{Ne}{ix} & i [em.] & 9.23 & \citetalias{Hanke09}\\
		\hline
		9.37   & 3.3 & \ion{Ne}{x} & Ly $\epsilon$ & 9.36 & \citetalias{Hanke09} \\
		\hline
		\hline
		\vspace{0.5em} 
		\end{tabular}
		\hspace{0.5em}
		\begin{tabular}{c c l l c c}
		\hline
		\hline
		Spike [\AA] & $\sigma$ & Ion & Line(s) & Wl. [\AA] & Ref. \\ 
		\hline
		9.48   & 2.1 & \ion{Fe}{xxvi} & 2s $\rightarrow$ 3p & 9.536 & 1\\
		9.50   & 3.4 & \ion{Fe}{xxvi} & 2p $\rightarrow$ 3d & 9.532 & 1\\
		&& \ion{Mg}{i} & K$\beta$ & 9.521 & \citetalias{Bearden67} \\ 
		&& \ion{Ne}{x} & Ly $\delta$ & 9.481 & \citetalias{Garcia65}, \citetalias{Hanke09} \\ 
		\hline 
		10.01  & 2.1 & \ion{Fe}{xvii} & 2s$^2$2p$^6$ $\rightarrow$ 2s2p$^6$5p & 10.102 &  1\\
		&& \ion{Fe}{xxv} & 2s $\rightarrow$ 3p & 10.037 &  1\\
		&& \ion{Na}{xi} & Ly $\alpha$ & 10.025 & \citetalias{Johnson85}, \citetalias{Hanke09}\\ 
		\hline
		10.63  & 2.2 & \ion{Fe}{xix} & 2p$^4$ $\rightarrow$ 2p$^3$4d & 10.664 & 1\\
		&&\ion{Fe}{xxiv} & 2s $\rightarrow$ 3p & 10.663 & 1\\
		&&\ion{Fe}{xvii} & 2p$^6$ $\rightarrow$ 2p$^5$6d & 10.658 & 1\\
		&&\ion{Ne}{ix} & He $\epsilon$ & 10.643 & \citetalias{Verner96}, \citetalias{Hanke09}\\ 
		&&\ion{Fe}{xxiv} & 2s $\rightarrow$ 3p & 10.619 & 1\\
		\hline 
		11.57  & 2.8 & \ion{Na}{i} & K$\beta$ & 11.575 & \citetalias{Bearden67} \\ 
		&&\ion{Ne}{ix} & He $\beta$ & 11.546 & \citetalias{Liao13}, \citetalias{Hanke09} \\ 
		&&\ion{Fe}{xviii} & 2p$^5$ $\rightarrow$ 2p$^4$4d & 11.539 & 1 \\
		&&\ion{Fe}{xviii} & 2p$^5$ $\rightarrow$ 2p$^4$4d & 11.524 & 1 \\
		\hline
		12.47  & 2.1 & \ion{Fe}{xx} & 2p$^4$ $\rightarrow$ 2p$^3$3d & 12.475 & 1 \\
		&& \ion{Fe}{xxi} & 2p$^2$ $\rightarrow$ 2p3d & 12.445 & 1\\
		\hline
		12.57  & 2.2 & \ion{Fe}{xx} & 2s$^2$ 2p$^3$ $\rightarrow$ 2s 2p$^3$ 3p & 12.576 &  1\\ 
		&& \ion{Fe}{xxi} & 2p$^3$ $\rightarrow$ 2p$^2$3d & 12.602 & 1\\
		&& \ion{Fe}{xx} & 2p$^4$ $\rightarrow$ 2p$^3$3d & 12.592 & 1\\
		&& \ion{Fe}{xx} & 2s$^2$2p$^3$ $\rightarrow$ 2s2p$^3$3p & 12.570 & 1\\
		&& \ion{Fe}{xx} & 2s$^2$2p$^3$ $\rightarrow$ 2s2p$^3$3p & 12.569 & 1\\
		&& \ion{Fe}{xx} & 2s$^2$2p$^3$ $\rightarrow$ 2s2p$^3$3p & 12.558 & 1\\
		\hline
		12.70  & 2.3 & \ion{Fe}{xxii} & 2p$^2$ $\rightarrow$ 2p3s & 12.710 & 1\\
		&&\ion{Fe}{xvii} & 2p$^6$ $\rightarrow$ 2p$^5$4s & 12.702 & 1\\
		\hline
		12.85  & 3.2 & \ion{Fe}{xx} & 2p$^3$ $\rightarrow$ 2p$^2$3d & 12.867 & 1\\
		&& \ion{Fe}{xx} & 2p$^3$ $\rightarrow$ 2p$^2$3d & 12.809 & 1\\
		&& \ion{Fe}{xviii} & 2s$^2$2p$^5$ $\rightarrow$ 2s2p$^5$3p & 12.801 & 1\\
		\hline
		\hline
		\vspace{0.5em} 
	\end{tabular}
	\caption*{\textbf{References}: (1) AtomDB (\url{http://www.atomdb.org/}), \citepalias{Bearden67} \citealt{Bearden67}, \citepalias{Drake88} \citealt{Drake88}, \citepalias{Garcia65} \citealt{Garcia65}, \citepalias{Hell16} \citealt{Hell16}, \citepalias{Johnson85} \citealt{Johnson85}, \citepalias{Liao13} \citealt{Liao13}, \citepalias{Verner96} \citealt{Verner96}. \textbf{Notes}: \citepalias{Hanke09} obtain line energies from (1, \citetalias{Verner96}). \citepalias{Hell16} measure blended line complexes ("L-shell" for short in the table). Lines listed multiple times with different wavelengths differ in the fine structure. Emission lines are marked with [em.]. Iron lines are exclusively from AtomDB.}
	\label{tab:spikes_3814}
\end{table*}

Figure~\ref{fig:rms} shows the \fvsq\ spectrum for both investigated data sets, Obs.~3814 and 11044, taken during the superior and inferior conjunction passage, respectively, in the $2\mbox{--}14$\AA\ range. A major difference exists between the two: at inferior conjunction, the variability is consistent with zero, while a continuum is present at superior conjunction, which linearly increases with wavelength above ${\sim}3$\AA. The linear fit shown in Fig.~\ref{fig:rms} serves to qualitatively characterise this trend and the significance of the superimposed features, the most prominent of which is a spike-like increase in variability just below 7\AA. In contrast, the observation at inferior conjunction shows no clear features. 

Figure~\ref{fig:3814} shows the \fvsq\ spectrum at superior conjunction for a higher resolution, relative to the linear fit. All data points that lie more than $2\sigma$ above a piece-wise running average over 101 bins are highlighted and listed in Tab.~\ref{tab:spikes_3814} with associated transitions. The clear feature seen in Fig.~\ref{fig:rms} just below 7\AA\ corresponds to the silicon line region. At the increased resolution, the feature splits into multiple spikes at positions corresponding to those of the lines. A similar group of spikes is seen in the sulphur region, but is less prominent, which is likely due to a reduction in signal caused by gap filtering strongly increasing the uncertainty (see Sect.~\ref{sec:sysvar} and Appendix~\ref{sec:gapfiltering}). An alignment of lines and spike features exists outside the silicon and sulphur regions as well: hints of neon (e.g., \ion{Ne}{x} Ly $\delta\mbox{--}\eta$ at $9.2\mbox{--}9.5$\AA), magnesium (e.g., \ion{Mg}{xii} Ly $\gamma$ at 6.74\AA), aluminium (\ion{Al}{xii} He $\beta$ at 6.64\AA), and iron (e.g.\ at 12.4--12.9\AA) features can be seen in Fig.~\ref{fig:3814}. In general, more spikes seem to be present close to known lines than in line-free regions (see black and blue in Fig.~\ref{fig:3814}). However, due to the low signal to noise ratio and strength of the \fvsq\ features, such detections are tentative. As previously mentioned, the \fvsq\ spectrum of Obs.~11044 shows far fewer features. A higher resolution \fvsq\ spectrum can be found in Fig.~\ref{fig:11044} in the appendix.

In Sect.~\ref{sec:s_si}, we will discuss the silicon and sulphur line regions in Obs.~3814 in more detail, as these contain the most prominent features. Their time dependence will be investigated in Sect.~\ref{sec:timedep}. As this work is among the first studying \fvsq\ spectra at gratings resolution \citep[predated only by][to our knowledge]{Mizumoto17}, we aim to assess the prospects of this approach and do not perform a rigorous search and identification of spike features. In addition, a quantitative analysis is challenging, because it requires an accurate description of the continuum. 

\subsection{Silicon and sulphur line regions at superior conjunction}
\label{sec:s_si}

\begin{figure*}
    \centering
    \includegraphics[width=0.48\textwidth]{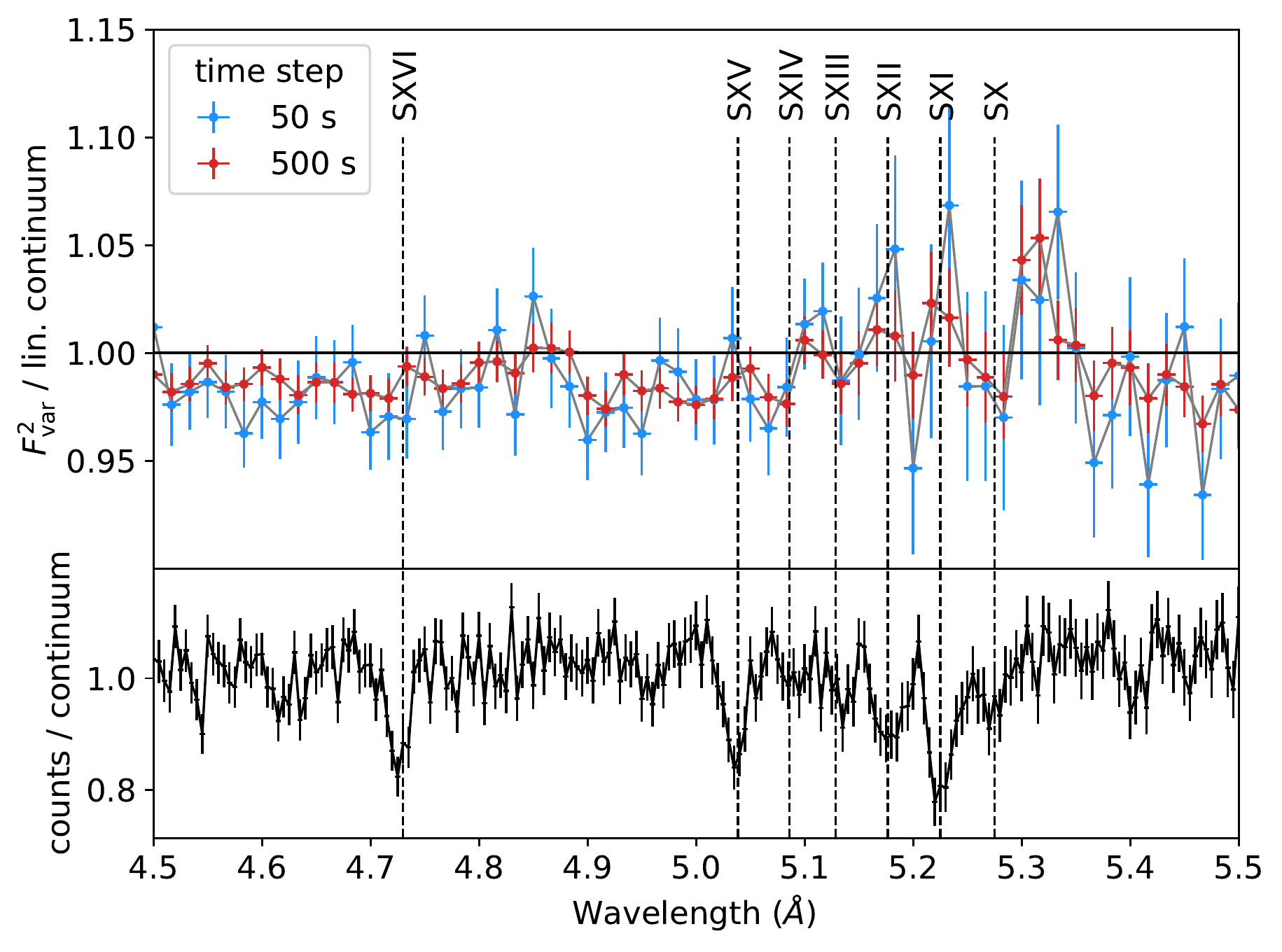}
    \includegraphics[width=0.48\textwidth]{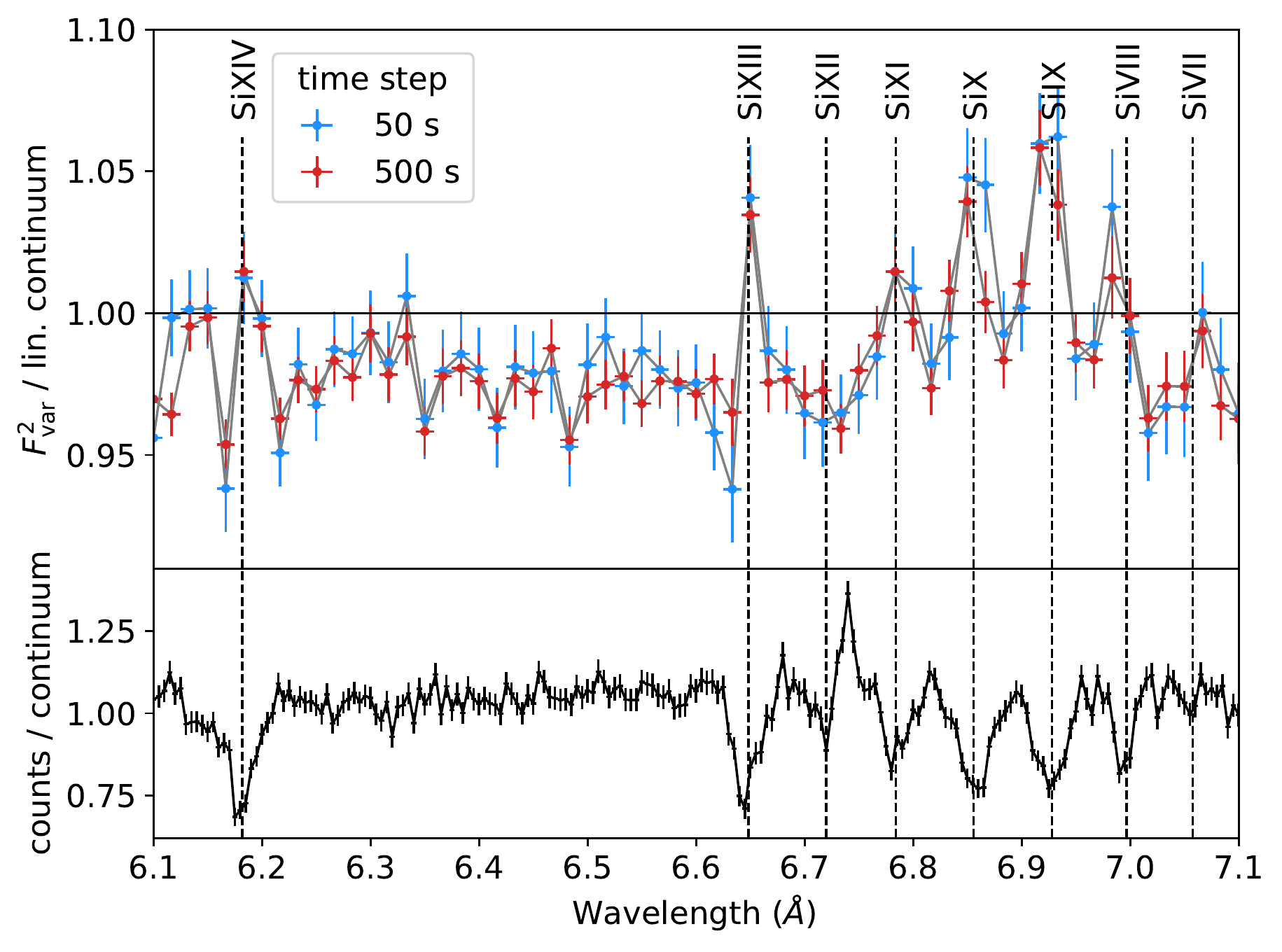}
    \caption{Sulphur (left) and silicon (right) line regions. The \fvsq\ spectrum of Obs.~3814 is shown relative to the linear fit from Fig.~\ref{fig:rms} (black) and compared to the count spectrum (lower panels) from \cite{Hirsch19} and the rest-wavelength of detected lines. The count spectrum is given relative to a local powerlaw fit, in the wavelength bands indicated in the figure, and is the sum of spectra of all dip stages. Two time binnings are displayed for the \fvsq\ spectrum, $50\,$s (blue) and $500\,$s (red), to investigate the effect of the added higher frequencies. Lines connect the data points for clarity. Overall, there is a good correspondence between \fvsq\ spike features and known line positions. }
    \label{fig:S_Si}
\end{figure*}

Figure~\ref{fig:S_Si} shows a zoom-in on the silicon and sulphur regions. In addition to the 500$\,$s binning investigated before, a 50$\,$s binning is shown to investigate if there is additional variability at higher frequencies. The bottom panels of Fig.~\ref{fig:S_Si} show the count spectra from \cite{Hirsch19}, summed for all dip stages, relative to local powerlaw fits in the given bands, to allow for a direct visual comparison to the \fvsq\ spectrum.  

For silicon, a spike is clearly detected for each corresponding line, except \ion{Si}{xii}, which might be because it cancels with the neighbouring He-like Si forbidden emission line. Line emission typically decreases variability, because, when looking at an object, the observer sees the sum of emission from all regions, in which variability in individual regions is averaged out \citep{Parker17_iras}. Visually, the spike positions and strengths directly correspond to those of the lines. The asymmetries present in some of the absorption lines in the analyses of the count spectra by \citet{Miskovicova16} and \citet{Hirsch19} are also hinted at in the spikes, albeit with greater uncertainties and lower resolution. The heights and widths of the spikes are slightly larger in the 50$\,$s binning, but this increase is not enough to exceed the uncertainty. 

The analysis of the sulphur region is complicated by greatly reduced signal, as laid out in Sect.~\ref{sec:res_overview}. Nevertheless, there is a small but visible increase in overall variability. Spike features are resolved and approximately coincide with line positions, with the best agreement for \ion{S}{xi} and \ion{S}{xii}. \ion{S}{xiii} and \ion{S}{xiv} cannot be resolved in the \fvsq\ spectrum, resulting in a joined spike. A spike feature with no corresponding line is present at ${\sim}5.33$\AA, but does not challenge the general conclusion that spike and line positions coincide, due to the large uncertainties and the clear result for silicon. As for silicon, \fvsq\ slightly increases for the shorter, $50\,$s time step, but remains within the uncertainty.

\subsection{Frequency dependence at superior conjunction}
\label{sec:timedep}

\begin{figure}
    \centering
    \includegraphics[width = \linewidth]{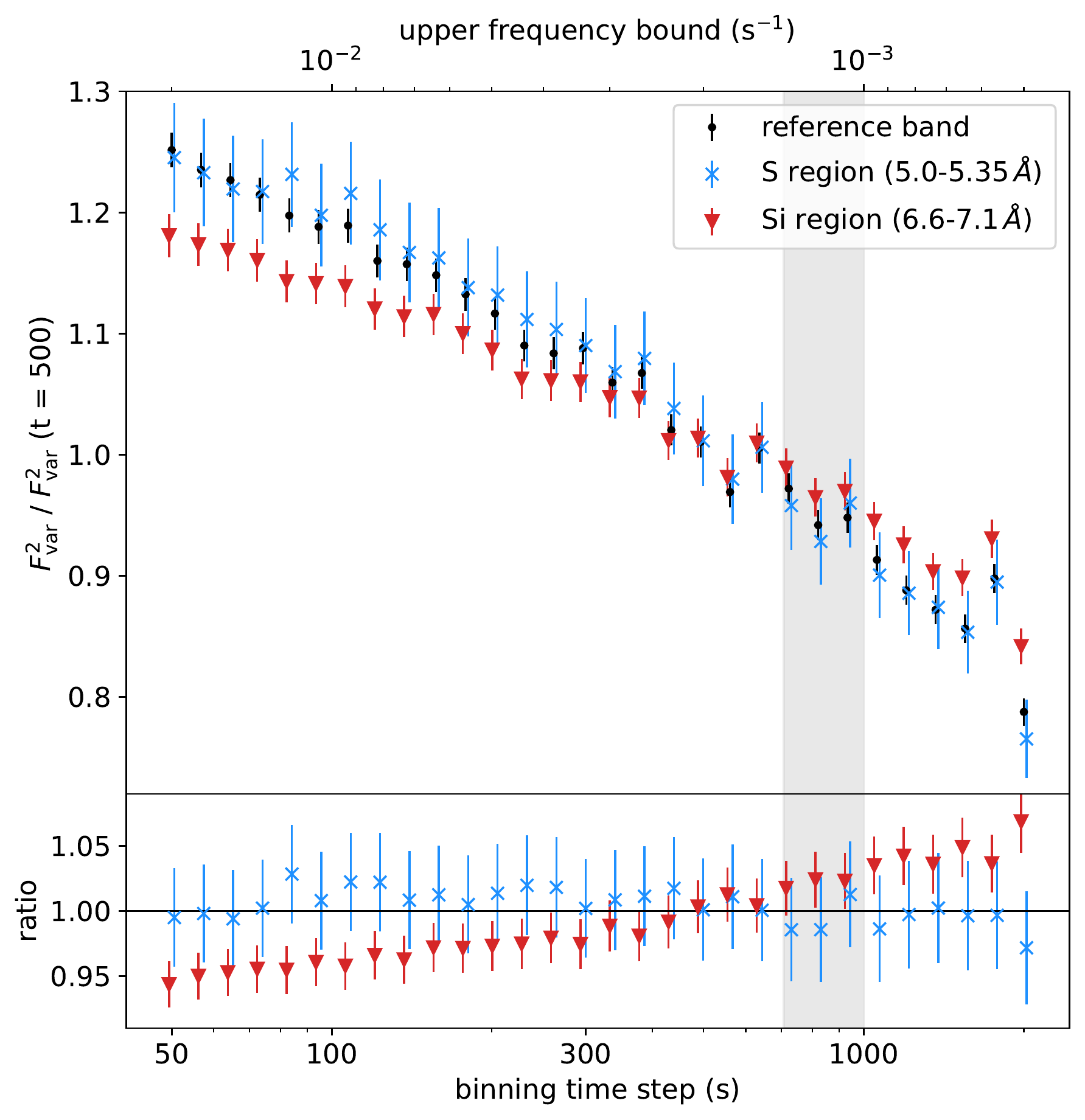}
    \caption{Total variability in Obs.~3814 in terms of \fvsq\ as a function of the binning time step, i.e., the figure effectively shows integrated frequency power. The sulphur (blue, $5\mbox{--}5.35$\AA) and silicon (red, $6.6\mbox{--}7.2$\AA) line regions are compared to a reference band (black, $5.4\mbox{--}6.6$\AA, excluding a region of $\pm 0.05$\AA\ around the \ion{Si}{xiv} line), with the ratio shown in the bottom panel. \fvsq\ is normalised to its value at $500\,$s in each respective band. The light grey band indicates a range where systematic effects due to the dither of the telescope would arise, if they were present.}
    \label{fig:timestep}
\end{figure}

The timescale of the variability can give an insight into its producing mechanism, for example the typical sizes of overdensities in a stellar wind that produces variability by obscuration (e.g., EG20). We investigate this timescale by changing the upper bound of the frequency range \fvsq\ integrates over, while keeping the lower bound fixed at the length of the observation. In practice, the upper bound is set by the binning time step of the light curves from which \fvsq\ is calculated (see Sect.~\ref{sec:fvsq}). The result in Fig.~\ref{fig:timestep} therefore effectively shows an integrated frequency spectrum, that is to say, its slope indicates the variability power added to \fvsq\ at any given frequency.  

In the figure, bands containing the bulk of detected silicon ($6.6\mbox{--}7.2$\AA) and sulphur ($5\mbox{--}5.35$\AA) lines are compared to a reference band, spanning the range in between them ($5.4\mbox{--}6.6$\AA), but excluding a region of $\pm 0.05$\AA\ around the \ion{Si}{xiv} line. The bottom panel shows the ratio of \fvsq\ in the line regions to the reference band. The \fvsq\ in each band is rescaled to its value at $t=500\,$s: $0.137\pm0.001$, $0.105\pm 0.002$, and $0.082\pm0.004$, for the silicon, sulphur, and the reference band, respectively. Note that these values are not normalised to the linear continuum fit, which means that \fvsq\ is highest in the silicon region due to the continuum, not the line variability. For the choice of time steps, refer to Sect.~\ref{sec:obs}. 

Overall, \fvsq\ declines approximately linearly with binning time step in the logarithmic scaling, which is consistent with an underlying red noise process. The silicon line region shows a flatter trend than the reference band, in other words, the power is distributed more towards lower frequencies in this region. For sulphur, the behaviour is consistent with the continuum, but subject to large uncertainties. 

A possible systematic effect is the periodicity of the dither in the telescope pointing. The light grey band in Fig.~\ref{fig:timestep} indicates the range where such an effect would be expected\footnote{https://cxc.cfa.harvard.edu/ciao/why/dither.html (\chandra\ CIAO supplemental information on dither)}. As no significant increase in \fvsq\ relative to the overall trend is present in this region, we conclude that the dither periodicity is not an issue in our analysis.

\section{Discussion}
\label{sec:discussion}

\subsection{Clump driven variability}
\label{sec:d_mechanism}

Variability can be caused by intrinsic changes in flux, originating close to the compact object, or be a result of obscuration by material passing through the line of sight. In the latter case, the variability can be greatly sensitive to the line of sight. We observe that while the continuum variability at superior conjunction increases linearly with wavelength, it is constant and consistent with zero at inferior conjunction. Additionally, while the light curve is constant at inferior conjunction, strong transient dipping events are present at superior conjunction \citep{Miskovicova16}. It is well established (cf.\ Sect.~\ref{sec:cyg_clumps}) that the dipping events are caused by clumps in the stellar wind. During both observations, \src\ was in the low/hard state at very similar fluxes. Jointly, these results strongly support the conclusion that the observed variability is driven by clumps crossing the line of sight.

On the continuum, spikes of increased variability are observed, with a clear correspondence between their properties, such as their position and strength, and those of absorption lines in the silicon line region and also, at reduced significance due to lower sensitivity, in the sulphur line region (cf.\ Sect.~\ref{sec:s_si}). This result suggests that the spikes are caused by an increased variability inherent to the absorption lines, which is also observed in ultra-fast outflows of AGN \citep[e.g.,][]{Parker17_nature, Parker18_PCA, Igo20}. A general prevalence of spikes near known lines outside the silicon and sulphur regions further supports this conclusion. Both intrinsic and absorption variability can cause enhanced \fvsq\ in lines. In ultra-fast outflows, the enhancement is consistent with the ionisation responding to intrinsic luminosity changes \citep[e.g.,][]{Pinto18, Parker20}. Considering again that \src\ has a clumpy wind and that the clumps have a layered ionisation structure (see Sect.~\ref{sec:cyg_clumps}), it seems plausible that ionisation could also be driving the variability in \src. However, in an optically thin medium such as the wind, the line strength can directly respond to a change in the column density, without requiring an associated ionisation change. It is likely that both mechanisms contribute to the observed enhancement of the line variability. The role of ionisation is discussed further in Sect.~\ref{sec:d_ioni}.

\subsection{Modelling the orbital phase dependent variability of a clumpy stellar wind}
\label{sec:d_sim}

\begin{figure}
    \centering
    \includegraphics[width = \linewidth]{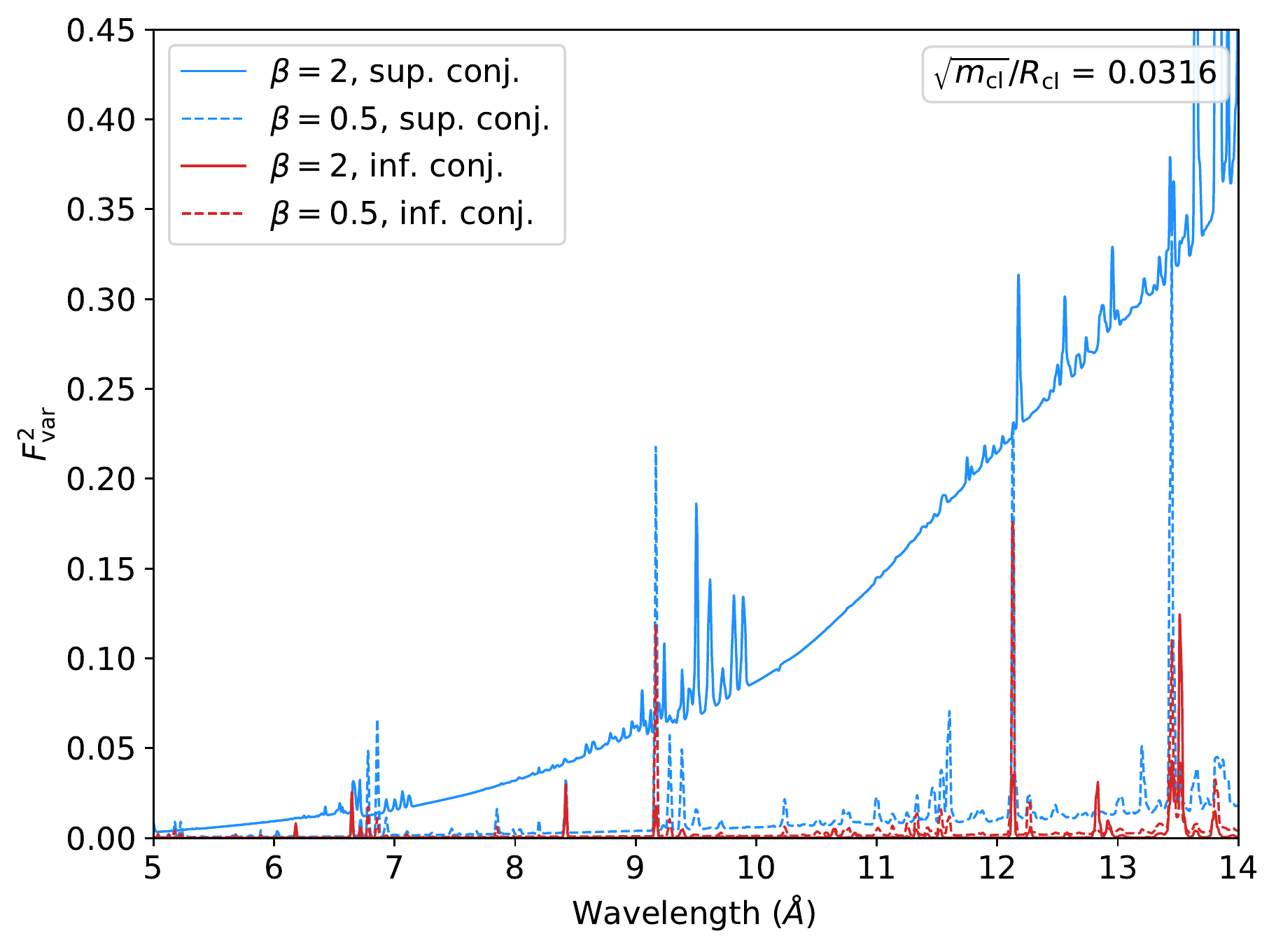}
    \includegraphics[width = \linewidth]{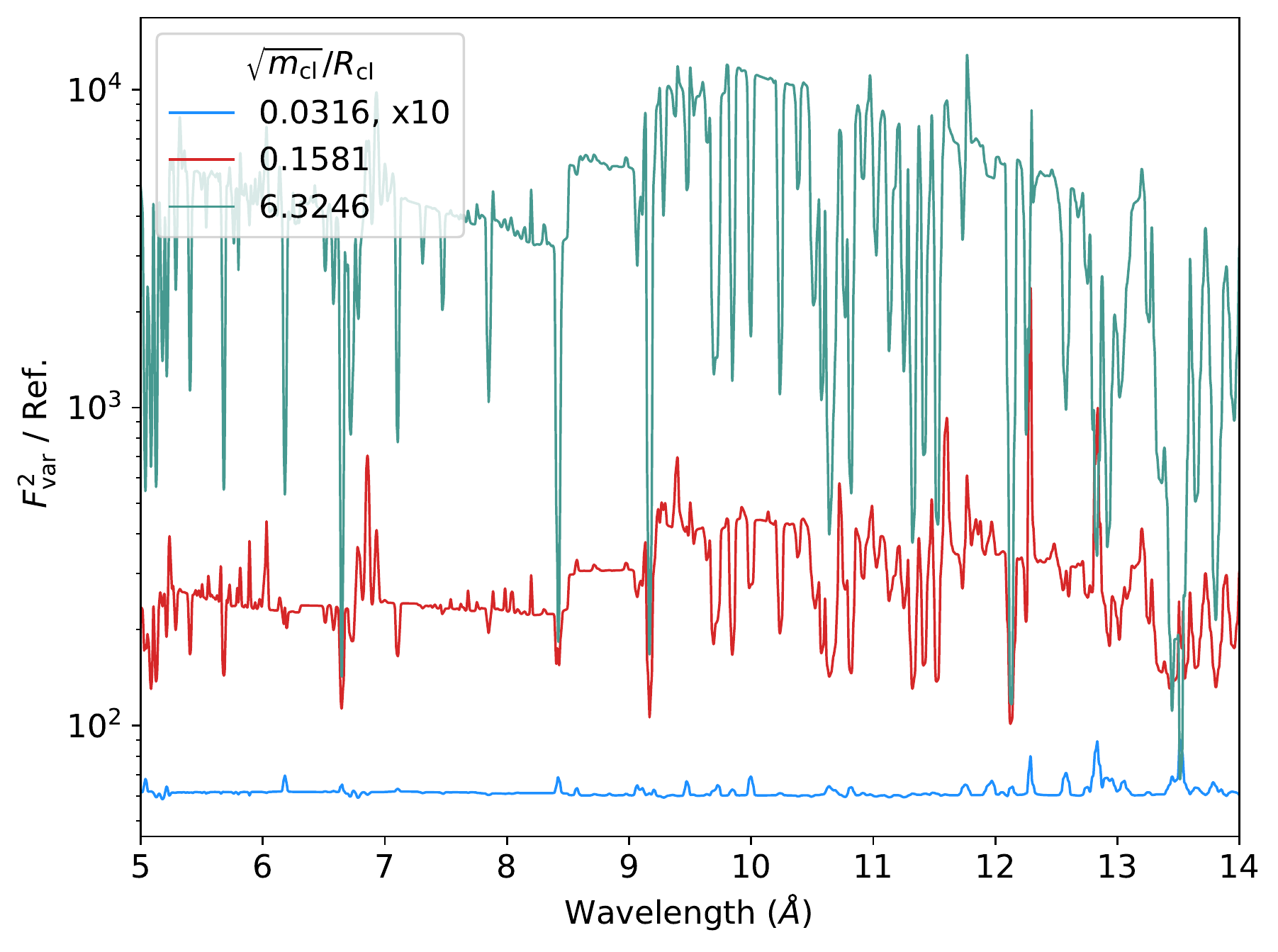}
    \includegraphics[width = \linewidth]{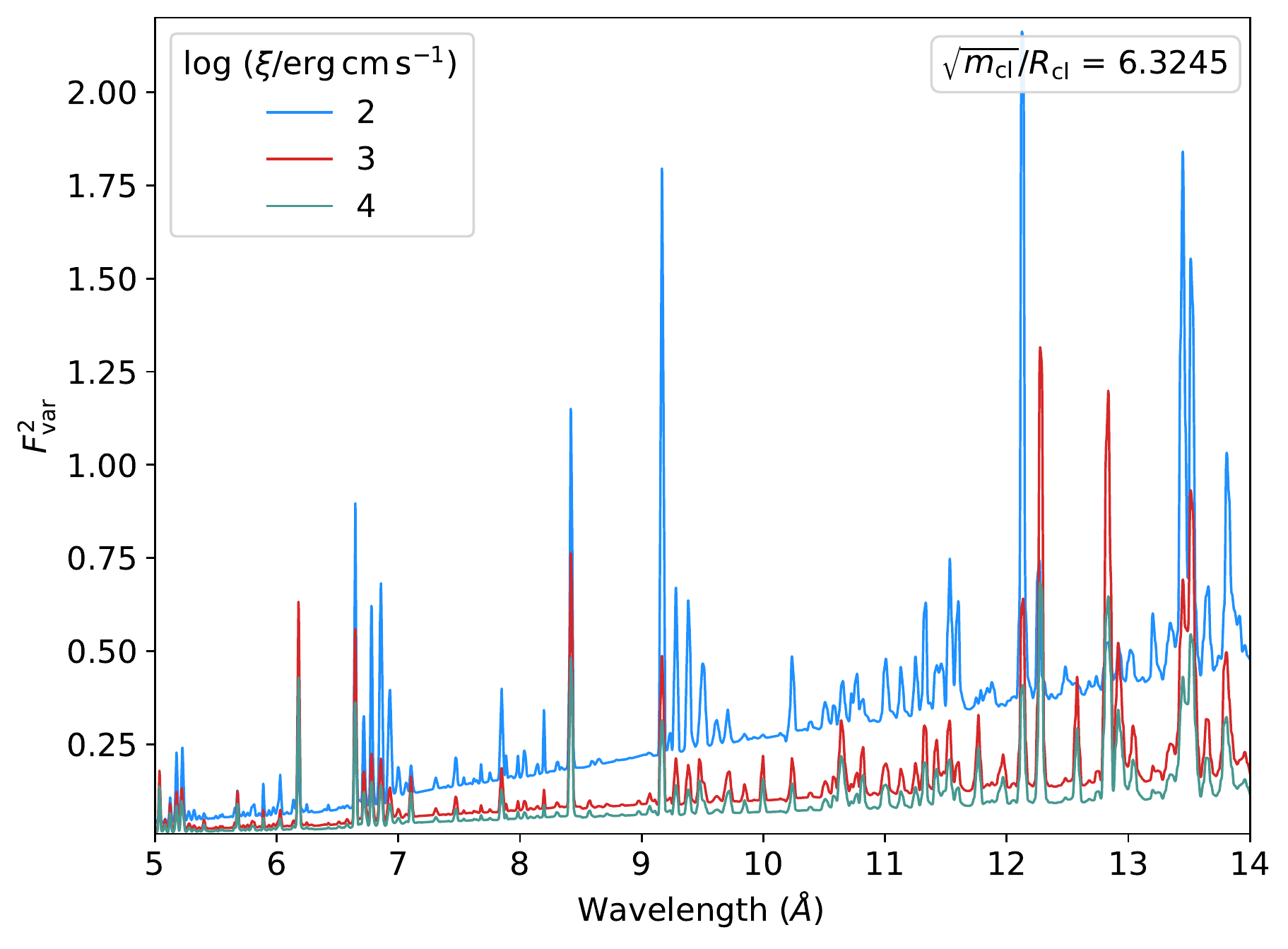}
    \caption{\fvsq\ expected for absorption in a clumpy wind following EG20. Details on the model generation can be found in the text. Upper panel: dependence on the line of sight (blue: superior conjunction, red: inferior conjunction) and the $\beta$ value of the wind velocity law (Eq.~\ref{equ:windvel}; solid: $\beta = 2$, dashed: $\beta = 0.5$), highlighting their strong impact on the continuum. Middle panel: ratio to a baseline model with $m^{0.5}_\mathrm{cl}R_\mathrm{cl}^{-1} = 0.0079$, showing the increase in continuum variability and with $m^{0.5}_\mathrm{cl}R_\mathrm{cl}^{-1}$ and the reduced increase of the line variability (dips in the ratio). The lowest-ratio model (blue) has been multiplied by a factor of 10 for better visibility. Bottom panel: ionisation dependence. A high $m^{0.5}_\mathrm{cl}R_\mathrm{cl}^{-1} = 6.3245$ is shown to highlight the continuum changes. The increase in \fvsq\ with ionisation is weaker for higher ionisations. For clarity, the model is only shown above 5$\AA$. }
    \label{fig:sim}
\end{figure}

In the last section, clumps passing through the line of sight were identified as main driver of the variability. We now compare the results of the \fvsq\ analysis to a model based on a prescription for a clumpy stellar by EG20. This model does not take into account X-ray photoionisation effects, but instead calculates the total column density of material along the line of sight. We therefore approximate the ionisation effects using an \textsc{xstar} \citep{Kallman01} table model as described below. At the moment, no model exists that simultaneously incorporates both clumping and ionisation.

EG20 model the X-ray absorption by a clumpy wind in a binary system consisting of a massive star and a compact object, described as an X-ray point source. The mass is assumed to be contained in spherical clumps, with a radius $R_\mathrm{cl}$ at two stellar radii, $2R_\star$, from the stellar centre, growing in size and accelerating as they move outwards according to the canonical velocity law for stellar winds, 
\begin{equation}
    \label{equ:windvel}
    v(r)=v_\infty(1-R_\star/r)^\beta \, ,
\end{equation}
where $\beta$ is a factor determining how quickly the terminal wind velocity, $v_\infty$, is reached. We employ the smooth expansion law for the clump size (Eq.~4 in EG20) but note that the impact of the clump expansion law on the variability is expected to be small (see EG20). The wind is assumed to be spherically symmetric and fast enough to be unaffected by the gravity of the compact object, which is a valid assumption for \src\ for two reasons: the terminal wind speed ($2100\,\mathrm{km}\,\mathrm{s}^{-1}$) is high compared to the orbital speed ($500\,\mathrm{km}\,\mathrm{s}^{-1}$) and more importantly, the bulk of material intercepting the line of sight is above the orbital plane.

The model calculates the total equivalent hydrogen column density, $N_\mathrm{H}$, resulting from the clumpy wind for a given line of sight parameterised by the inclination, $i$, and orbital phase, $\varphi$, without taking into account possible ionisation effects. We adopt the parameters that EG20 give for \src: $i = 27.1^\circ$, mass outflow rate, $\dot{M} = 3\cdot 10^{-6}\,\mathrm{M}_\odot\,\mathrm{yr}^{-1}$, $v_\infty = 2100 \, \mathrm{km}\,\mathrm{s}^{-1}$, and orbital separation, $a = 2.5\,R_\star$. Orbital phases of $\pm0.1$ around superior and inferior conjunction are covered, with a resolution of 20000 points, resulting in a time step of ${\sim} 5\,$s. For further details on the model and its application to \src\ see EG20. 

The $N_\mathrm{H}$ light curve obtained from the clumpy wind model is then used to create absorbed model spectra, from which \fvsq\ can be calculated. We do this by faking an absorbed powerlaw for each value in the $N_\mathrm{H}$ light curve with the \textsc{isis} command \texttt{fakeit}. The powerlaw index and normalisation were fixed to the values obtained by \citet{Hanke09} for the \chandra\ and \textit{RXTE} joined continuum of Obs.~3814 ($\Gamma=1.60$, norm = 1.33, see Tab.~2 in the reference). For the absorption, we use a custom \textsc{xstar} table model, following the approach detailed in \citet{Hirsch19}. The model is based on a calculation of the ionisation balance in a spherical gas cloud irradiated from the centre. The SED describing the source of incident continuum at the centre of the cloud includes a \texttt{nthcomp} continuum model, combined with lamp-post model relativistic reflection (\texttt{relxilllpcp}) and a small contribution from a non-relativistic reflection (\texttt{xillvercp}). The model reflects the idea that the primary X-ray radiation of the BH is already reprocessed by reflection of the accretion disk before interacting with the stellar wind of the donor. For this study, we chose $\Gamma=1.67$, $kT_\mathrm{e}=400\,\mathrm{keV}$, and inclination of $36^\circ$\footnote{A misalignment of the inner disk inclination to the inclination of the system has been inferred for \src\ in several studies when not using high density reflection models \citep{Tomsick_2014a,Parker_2015a,Walton_2016a,Tomsick_2018a} and is also supported by polarimetry measurements \citep{Krawczynski22}. The value we employ here corresponds to the measurements of inner disk inclination in the hard state in \citet{Tomsick_2018a}.} for both reflection components. The height of the primary component in the \texttt{relxilllpcp} component is $6.1\,r_\mathrm{g}$ and the spin of the BH is $0.998$. The ionization parameters $\log\xi$ are 3.1 and 0, for \texttt{relxilllpcp} and  \texttt{xillvercp}, respectively, and the reflection fractions are $0.27$ and $-1$\footnote{See \url{http://www.sternwarte.uni-erlangen.de/~dauser/research/relxill/} for the definition of the parameters.}. The reflection parameters represent a reasonable assumption for the conditions in \src\ in the typical hard state. The results of our qualitative discussion, which is concerned with the overall effects of ionisation, are not sensitive to slight changes in the conditions.

The \fvsq\ spectra resulting from this approach are shown in Fig.~\ref{fig:sim}. We investigate the dependence on $\beta$, the line of sight, the ionisation parameter of the \textsc{xstar} model\footnote{As defined by \citet[][]{Tarter69}: $\xi = L_\mathrm{X} \rho^{-1} r^{-2}$\,.}, $\xi$, and the ratio $m_\mathrm{cl}^{0.5}R_\mathrm{cl}^{-1}$, where $m_\mathrm{cl}$ and $R_\mathrm{cl}$ are given as multiples of $m_0 = 1.1\cdot 10^{18}\,\mathrm{g}$ and $R_\star = 17\,\mathrm{R}_\odot$\footnote{Note that we adopt the stellar radius given in EG20 here, instead of the updated value of $R_\star = 22.3\,\mathrm{R}_\odot$ \citep{Miller-Jones21}. The results are unaffected by this slight inconsistency, because $R_\star$ only functions as a scaling factor for the $m_\mathrm{cl}^{0.5}R_\mathrm{cl}^{-1}$ ratio.}. The clump mass, $m_\mathrm{cl}$, takes values of 0.4 and 10 and $R_\mathrm{cl}$ ranges over $0.005\mbox{--}0.08$. This range overlaps with the result of \citet{Feng_2002a} ($R_\mathrm{cl} = 0.01\mbox{--}0.6$), who obtained clump sizes by analysing the duration of the dips in the light curve. The choice of the combined parameter $m_\mathrm{cl}^{0.5}R_\mathrm{cl}^{-1}$ is motivated by EG20, who found that it is proportional to the spread of the column density, $\delta N_\mathrm{H}$. If variability is caused by $N_\mathrm{H}$ changes, as our model assumes, then $\delta N_\mathrm{H}$ directly translates in \fvsq\ variability. We consider two values for $\beta$, 0.5 and 2. The former is widely used in the literature and believed to be representative of winds of early O supergiants \citep{Sundqvist19}, while the latter corresponds to the velocity profiles computed by \cite{Sander18}, showing a more gradual acceleration. Unless otherwise indicated, the ionisation is fixed at $\log \xi = 3$, $\beta = 0.5$, and variability at inferior conjunction is shown. The ionisation, $\xi$, is always given in units of $\mathrm{erg}\,\mathrm{cm}\,\mathrm{s}^{-1}$.

\subsection{Line of sight dependence of the variability continuum}

The employed model is able to qualitatively reproduce the data, by predicting an approximately linear continuum and spikes of increased \fvsq\ (Fig.~\ref{fig:sim}, upper panel). The model also shows a strong line of sight effect, with higher variability at superior conjunction. The effect is especially strong for the higher $\beta$ value, $\beta = 2$, and is a direct consequence of how $\delta N_\mathrm{H}$ depends on the orbital phase. The clumps probed at superior conjunction are smaller, because the line of sight runs closer by the star than at inferior conjunction and the clumps expand as they move outwards. As $\delta N_\mathrm{H}\propto m_\mathrm{cl}^{0.5}R_\mathrm{cl}^{-1}$, the variability is higher at superior conjunction if the clump mass is fixed. A high $\beta$ increases this effect, because the slower acceleration requires a higher density if $\dot{M} = \mathrm{const.}$ and therefore more clumps. The variability increases with wavelength, because lower energy photons are 
more sensitive to changes in absorption.

For $\beta = 2$ and $m_\mathrm{cl}^{0.5}R_\mathrm{cl}^{-1} = 0.0316$, the model predicts $F_\mathrm{var}^2 \sim 0.37$ at 14\AA\ at superior conjunction and an inferior conjunction continuum that is consistent with zero within the measurement accuracy. These values agree very well with the measurement (Fig.~\ref{fig:rms}). However, the slope of the superior conjunction continuum is too steep, which causes \fvsq\ to be underestimated below 14\AA. Larger $m_\mathrm{cl}^{0.5}R_\mathrm{cl}^{-1}$ reduce this discrepancy, but require a non-zero inferior conjunction continuum.

The fact that the model is able to reproduce the qualitative behaviour of the data consistently at both conjunctions shows that varying absorption in an ionised clumpy wind is a viable explanation for the observed variability. Discrepancies, such as mentioned above, are expected given that the model is in an early state and we thus discuss two noteworthy model caveats and their effect on the variability continuum and line of sight dependence. First is the spherical wind assumption, which disregards the focused stream between star and black hole,
whose higher density could lead to an increase in variability at superior conjunction. At inferior conjunction, the line of sight passes through the bow shock trailing the black hole, which could disrupt the wind structure and smooth out the absorption variability. A similar effect is expected from interaction with the jet, through which the wind inevitably passes before crossing the inferior conjunction line of sight. \citet{Perucho12} showed that wind structures are usually destroyed in interactions with a jet. Second is the use of the canonical velocity law (Eq.~\ref{equ:windvel}), especially considering that the wind is ionised by the X-ray emission from the black hole, resulting in a change in  the populations of the driving lines. The X-ray luminosity in \src\ \citep[$L_\mathrm{x}\gtrsim10^{37}\,\mathrm{erg}\,\mathrm{s}^{-1}$,][]{Miskovicova16, Hirsch19} is in a regime where it could significantly inhibit the wind acceleration \citep{krticka18}, although the net effect of X-ray photoionising feedback is still uncertain \citep[see][for the insightful example of Vela X-1, another HMXB]{Sander18}. The inhibiting effect would be largest towards the black hole where the X-ray irradiation is strongest. It could therefore increase \fvsq\ at superior conjunction, as a slower wind must have a higher density if $\dot{M}$ is fixed. Even without an ionising source, \citet{Sander17} predict a slower onset of the acceleration, from a model that calculates a self-consistent hydrodynamic stratification of the wind.

\subsection{Suppressed line variability at inferior conjunction}
\label{sec:linevar_infconj}

The line variability predicted by the model is of a similar strength at superior and inferior conjunction (Fig.~\ref{fig:sim}, upper panel), yet the observations show line variability only at superior conjunction. This discrepancy is due to the wind moving and becomes apparent from the discussion of the count spectra in \citet{Miskovicova16}. While the superior conjunction spectrum shows absorption, mainly P-Cygni profiles are observed at inferior conjunction. \citet{Miskovicova16} ascribe this difference to the fact that the absorption at inferior conjunction is highly blue-shifted, because of the line of sight projection of the wind velocity, separating it from the ubiquitous emission component. Emission lines are suggested to reduce \fvsq\ instead of increasing it \citep{Parker17_nature}. It is likely that the variability of the absorption and emission components cancels out, given that the absorption component is weak at inferior conjunction and that our resolution is lower than that of the count spectra. In addition, narrow spikes appear smaller at lower resolution, because of the averaging effect. The resolution effect also applies at superior conjunction, and plausibly explains why the reported variability is slightly lower than in the model (compare the \fvsq\ spikes in Fig.~\ref{fig:rms} and Fig.~\ref{fig:sim}, upper panel). 

\subsection{Clump mass and radius}
\label{sec:clumps_m_R}

The middle panel of Fig.~\ref{fig:sim} shows the behaviour of our model with $m_\mathrm{cl}^{0.5}R_\mathrm{cl}^{-1}$, by taking a low value as a baseline (0.0079; big, light clumps: $m_\mathrm{cl} = 0.4\,m_0$, $R_\mathrm{cl} = 0.08\,R_\star$) and showing the ratio of models with higher values to it. As mentioned in Sect.~\ref{sec:d_sim}, $m_\mathrm{cl}^{0.5}R_\mathrm{cl}^{-1}$ is proportional to $\delta N_\mathrm{H}$ and therefore also to $F_\mathrm{var}$. For the investigated ratios, the variability spans several orders of magnitude, making $m_\mathrm{cl}^{0.5}R_\mathrm{cl}^{-1}$ an influential parameter. For low ratios, the increase in the lines is stronger than in the continuum (light blue curve), but as higher ratios are approached, the increase is reduced (dips in the red and green curves). This relative reduction in line variability is an averaging effect. As $m_\mathrm{cl}^{0.5}R_\mathrm{cl}^{-1} \propto \delta N_\mathrm{H}$, the absorption varies over a larger range for higher values of the ratio. Less absorbed spectra are brighter and therefore have a higher weight in the average that is contained in the definition of \fvsq\ (see Eq.~\ref{equ:exvar} and \ref{equ:fvar}). They also show weaker lines, because the line strength responds to the absorbing column. As a consequence, the line variability relative to the continuum is weaker for higher $\delta N_\mathrm{H}$, resulting in the dips in the model ratio. In addition to this effect, the continuum variability gets more linear, as seen by comparing the upper to the bottom panel in Fig.~\ref{fig:sim}, where a high $m_\mathrm{cl}^{0.5}R_\mathrm{cl}^{-1}$ ratio was chosen.

An independent approach to obtain information on $R_\mathrm{cl}$ is measuring the timescale of the variability, as small structures cross the line of sight more quickly than large ones, causing the timescale to depend on $R_\mathrm{cl}$. Our analysis of the time binning dependence in Sect.~\ref{sec:timedep} revealed that the distribution of variability power is shifted towards lower frequencies in the silicon line region compared to the continuum. Such a result is explained by the response of the line region to the column density changes driving the variability: the line depths vary across a wider range in bigger clumps, which enhances their variability and therefore flattens the frequency trend. This mechanism is likely complemented by the layered ionisation structure of the clumps. Bigger clumps allow for a lower ionisation in their centre (see Sect.~\ref{sec:cyg_clumps}), which also increases the range across which line depths vary. In addition, the larger ionisation range gives rise to more lines. This interpretation is in line with our conclusion from Sect.~\ref{sec:d_mechanism} that column density changes might drive ionisation variability. As a general conclusion, timescale measurements can independently reveal $m_\mathrm{cl}$ and $R_\mathrm{cl}$, if $m_\mathrm{cl}^{0.5}R_\mathrm{cl}^{-1}$ can be constrained. In principle, the clump size can also be obtained from the duration of dips in the light curve, as demonstrated, in \citet{Feng_2002a} and \citet{Ness12}. In case of \src, the necessity to account for the orbital speed, the progressive acceleration of the wind, and projection effects complicates this approach (see discussion in EG20).

\subsection{Ionisation}
\label{sec:d_ioni}

\begin{figure*}
    \centering
    \includegraphics[width = \textwidth]{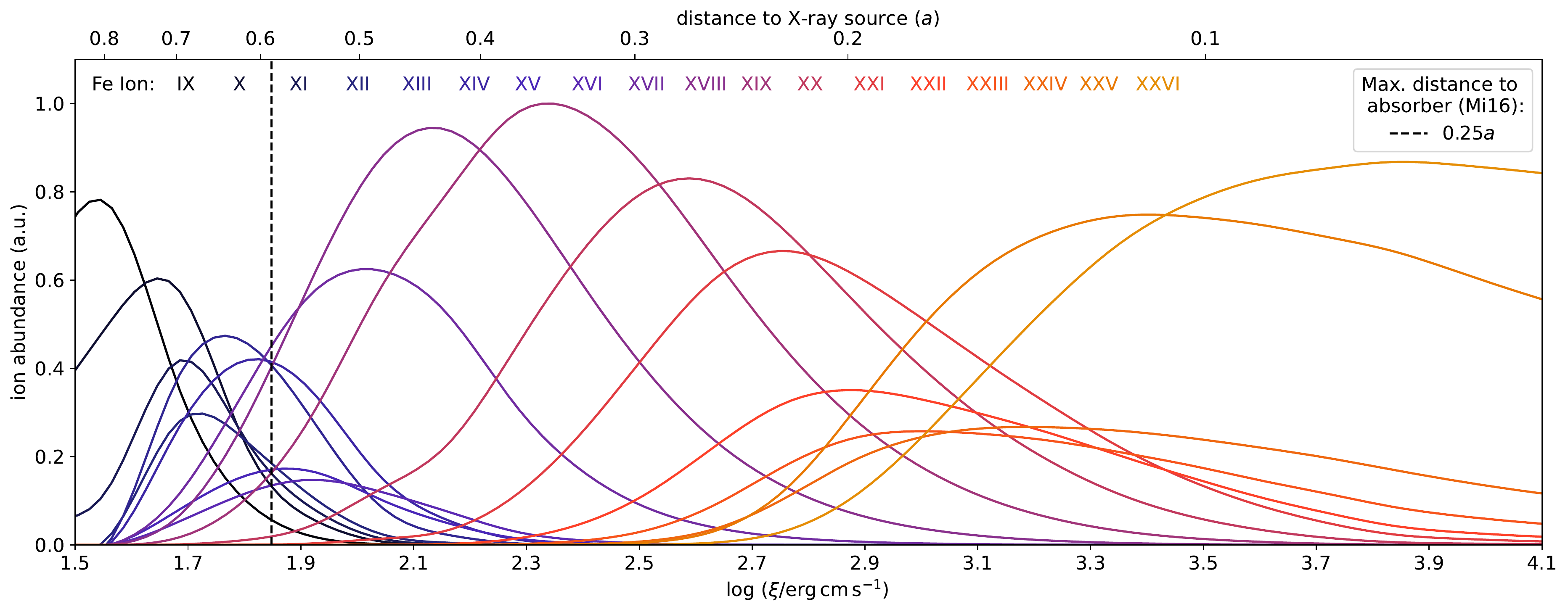}
    \caption{Relative abundances of iron K to M-shell ions as a function of the ionisation parameter, $\xi$ (lower x-axis) or, equivalently, the distance absorber--X-ray source, $d$ (upper x-axis). The dashed line marks the maximum value of $d$ according to \citet{Miskovicova16}, given in terms of the semi-major axis of the binary system, $a = 2.35\,R_\star$ \citep{Miller-Jones21}. For details, see Sect.~\ref{sec:d_ioni}.}
    \label{fig:fe-abund}
\end{figure*}

The bottom panel of Fig.~\ref{fig:sim} shows \fvsq\ model spectra for $\log \xi = 2\mbox{--}4$ (the $m_\mathrm{cl}^{0.5}R_\mathrm{cl}^{-1}$ ratio was chosen to be high in this plot to explicitly show the continuum changes.). Ionisation acts mainly as a scaling factor for both the continuum and line variability. The effect is stronger at lower ionisations. 

We therefore do not expect the choice of $\log \xi = 3$ to impact the conclusions we reached so far. To illustrate the effect of $\xi$ on spectral features in more detail, we show Fe ion abundances from an \textsc{xstar} photoionisation model in Fig.~\ref{fig:fe-abund}. The lower x-axis shows the maximum value of $\xi$ reached in the absorber. The corresponding minimum distance of the absorber from the X-ray source (upper x-axis) is obtained by inverting the definition of $\xi$ \citep{Tarter69},
\begin{equation}
    d = \sqrt{\frac{L_\mathrm{X}}{\xi\rho}} \, ,
\end{equation}
where $L_\mathrm{X} = 10^{37}\,\mathrm{erg}\,\mathrm{s}^{-1}$ for \src\ and $\rho$ is taken from the clumpy wind model by EG20 evaluated at superior conjunction (adopting $\beta = 2$ and a mean molecular weight of $\mu = 1\,$u). \citet{Miskovicova16} estimate the distance between absorber and black hole to be $d \lesssim 0.25\,a$, which gives $\log \xi \gtrsim 1.8$. 
This lower limit does not constrain the ion abundances well (see Fig.~\ref{fig:fe-abund}) and a conclusion on which lines dominate the spectrum is difficult to draw without better constraints on the absorber position. Considering that the absorber likely has a significant extend, is clumpy and that the ion abundances strongly depend on $\xi$ (see Fig.~\ref{fig:fe-abund}), it is expected to have a complex ionisation structure.

Despite the discussed limitations, fixing the ionisation to a single value has proven to be a successful first step, considering the good qualitative agreement of data and model. However, this simple approach inevitably results in an incomplete picture, because varying ionisation is known to contribute to the variability (see Sect.~\ref{sec:d_mechanism}) and even the clumps themselves have a layered ionisation structure. Examples for the behaviour that might arise from this structure were given in Sect.~\ref{sec:d_mechanism} and \ref{sec:clumps_m_R}. 

In the general case, the response of variability to changes in $\xi$ is non-trivial: for example, variability in absorption lines is not necessarily enhanced. The decisive question is how it correlates with the continuum variability. If line strength decreases with increasing flux, the total flux change is higher in the line than in the continuum, causing \fvsq\ to spike. In the contrary case, the change in the line counteracts that in the continuum, resulting in an \fvsq\ dip. For ionisation-driven variability, both cases are possible, even within the same observation because a change in ionisation shifts the ion balance, which means that abundances rise for some ions while they decrease for others. Only in specific regimes can the situation be simpler: for example, in an almost fully ionised environment, the responses of the lines to ionisation changes are more uniform. This special case is thought to apply to ultra-fast outflows of AGN and result in the observed anti-correlation between line strength and flux \citep{Parker17_nature, Parker18_PCA, Igo20}. In \src, the situation is likely to be complex, because of the wide range of ionisations. Adding to that, lines can show different responses in different regions in the source from which we observe the sum, as we argued in Sect.~\ref{sec:linevar_infconj} for P-Cygni profiles. As a result of this complex behaviour, a comprehensive model will require radiative transfer modelling, which is outside the scope of this work.

\section{Summary and Outlook}
\label{sec:concl}

We presented an analysis of high resolution excess variance spectra of \src\ calculated from \chandra/HETG observations during the low/hard state at superior and inferior conjunction (ObsID~3814 and 11044, respectively), which enabled us to discuss the impact of the accretion geometry and wind structure on the excess variance. We match features in the excess variance to their counterparts in the count spectra comparing to previous spectral analysis  by \citet{Hanke09}, \citet{Miskovicova16}, and \citet{Hirsch19} and to a selection of significant iron lines in the $2\mbox{--}14\AA$ range.
To complement the data analysis, we modelled excess variance spectra from X-ray absorption in a clumpy stellar wind based on EG20 and an \textsc{xstar} table model to incorporate ionisation effects. 

We have demonstrated that we can deal with systematic variability introduced by the use of \chandra\ gratings data, by accounting for photon loss in CCD gaps, differences in the effective areas of the gratings arms, the discreteness of the CCD read-out cycle, and the telescope dither. The fact that the inferior conjunction variability is consistent with zero and shows no features rules out the existence of further systematic effects that would increase the variability at a measurable level, highlighting the reliability of our results.  

Overall, this work demonstrates that excess variance can be used to study winds of HMXBs, providing an independent new approach to previous high resolution spectral studies.
Our main conclusions are:

\begin{enumerate}
    \item The excess variance is strongly dependent on the orbital phase with a linearly increasing continuum and spike features at superior conjunction and the inferior conjunction spectrum consistent with zero.
    
    \item Modelling shows that the variability is consistent with changes in column density caused by clumps crossing the line of sight. The orbital phase dependence is slightly under-predicted and a slowly accelerating wind is required. We suggest that the jet or the bow shock of the black hole in the focused wind disrupts the wind structure, suppressing inferior conjunction variability, and name effects that might increase superior conjunction variability.
    
    \item Variability in absorption lines is enhanced at superior conjunction, with this effect being especially clear in the silicon and sulphur line regions. We suggest that no significant variability spikes are seen at inferior conjunction, because the projected wind velocity is higher, causing the count spectrum to be dominated by P-Cygni profiles, whose absorption and emission components cancel out in variability.
    
    \item The variability power in the silicon region is redistributed towards lower frequencies, compared to the nearby continuum. We argue that the line variability might be increased in big clumps, which have higher column densities and reach lower ionisations in their centres. 
 \end{enumerate}

We have discussed that the impact of ionisation is highly non-tivial, calling for radiative transfer simulations. In future work, we aim to generate an \fvsq\ table model for clumpy stellar winds including improvements such as this one. Such a model will allow us to constrain the clump mass-size ratio, $m^{0.5}R_\mathrm{cl}^{-1}$, which constitutes a direct approach to quantify the wind structure, when combined with an \fvsq\ timescale analysis, the feasibility of which we have demonstrated. 

This outlook highlights that excess variance spectra are a versatile tool, first and foremost, because they manage to combine spectral and timing information, making it easy to handle and model. In addition, excess variance features tend to be more pronounced than those in count spectra (compare the height of the silicon \fvsq\ spike to the continuum in Fig.~\ref{fig:rms}), which is in line with previous results laid out in Sect.~\ref{sec:intro}. This emphasizes the importance of establishing this analysis method for the future missions. The results presented here also present a sneak peak at the detail with which extra-galactic sources will be able to be studied with microcalorimetry.
The major challenge, and the reason why few studies have applied excess variance spectroscopy to data of high spectral resolution, is the amount of signal required to perform simultaneous energy and time binning. Within the next decades, microcalorimetry instruments such as \xrism\ \citep{Tashiro18} and \athena\ \citep{Nandra13} will become available that solve this issue for a range of sources. The technical difficulties we faced in this work come almost exclusively from the use of gratings and will naturally disappear when switching to microcalorimetry.

\begin{acknowledgements}
      LH was supported by the European Space Agency (ESA) trainee programme. IEM has received funding from the European Research Council (ERC) under the European Union’s Horizon 2020 research and innovation programme (SPAWN ERC, grant agreement No 863412). RB acknowledges support by NASA under award number 80GSFC21M0002.
\end{acknowledgements}

%
%

\bibliography{aanda}

\begin{thebibliography}{74}
\expandafter\ifx\csname natexlab\endcsname\relax\def\natexlab#1{#1}\fi

\bibitem[{{Bearden}(1967)}]{Bearden67}
{Bearden}, J.~A. 1967, Reviews of Modern Physics, 39, 78

\bibitem[{{Boroson} \& {Vrtilek}(2010)}]{Boroson_2010a}
{Boroson}, B. \& {Vrtilek}, S.~D. 2010, \apj, 710, 197

\bibitem[{{Bowyer} {et~al.}(1965){Bowyer}, {Byram}, {Chubb}, \&
  {Friedman}}]{Bowyer65}
{Bowyer}, S., {Byram}, E.~T., {Chubb}, T.~A., \& {Friedman}, H. 1965, Annales
  d'Astrophysique, 28, 791

\bibitem[{{Brocksopp} {et~al.}(1999){Brocksopp}, {Fender}, {Larionov}, {Lyuty},
  {Tarasov}, {Pooley}, {Paciesas}, \& {Roche}}]{Brocksopp99}
{Brocksopp}, C., {Fender}, R.~P., {Larionov}, V., {et~al.} 1999, \mnras, 309,
  1063

\bibitem[{{Canizares} {et~al.}(2005){Canizares}, {Davis}, {Dewey}, {Flanagan},
  {Galton}, {Huenemoerder}, {Ishibashi}, {Markert}, {Marshall}, {McGuirk},
  {Schattenburg}, {Schulz}, {Smith}, \& {Wise}}]{Canizares05}
{Canizares}, C.~R., {Davis}, J.~E., {Dewey}, D., {et~al.} 2005, \pasp, 117,
  1144

\bibitem[{{Castor} {et~al.}(1975){Castor}, {Abbott}, \& {Klein}}]{Castor75}
{Castor}, J.~I., {Abbott}, D.~C., \& {Klein}, R.~I. 1975, \apj, 195, 157

\bibitem[{{Drake}(1988)}]{Drake88}
{Drake}, G.~W. 1988, Canadian Journal of Physics, 66, 586

\bibitem[{{Edelson} {et~al.}(2002){Edelson}, {Turner}, {Pounds}, {Vaughan},
  {Markowitz}, {Marshall}, {Dobbie}, \& {Warwick}}]{Edelson02}
{Edelson}, R., {Turner}, T.~J., {Pounds}, K., {et~al.} 2002, \apj, 568, 610

\bibitem[{{El Mellah} {et~al.}(2020){El Mellah}, {Grinberg}, {Sundqvist},
  {Driessen}, \& {Leutenegger}}]{ElMellah20}
{El Mellah}, I., {Grinberg}, V., {Sundqvist}, J.~O., {Driessen}, F.~A., \&
  {Leutenegger}, M.~A. 2020, \aap, 643, A9

\bibitem[{{El Mellah} {et~al.}(2018){El Mellah}, {Sundqvist}, \&
  {Keppens}}]{ElMellah18}
{El Mellah}, I., {Sundqvist}, J.~O., \& {Keppens}, R. 2018, \mnras, 475, 3240

\bibitem[{{Feng} \& {Cui}(2002)}]{Feng_2002a}
{Feng}, Y.~X. \& {Cui}, W. 2002, \apj, 564, 953

\bibitem[{{Friend} \& {Castor}(1982)}]{Friend82}
{Friend}, D.~B. \& {Castor}, J.~I. 1982, \apj, 261, 293

\bibitem[{{Fullerton} {et~al.}(2006){Fullerton}, {Massa}, \&
  {Prinja}}]{Fullerton06}
{Fullerton}, A.~W., {Massa}, D.~L., \& {Prinja}, R.~K. 2006, \apj, 637, 1025

\bibitem[{{Garcia} \& {Mack}(1965)}]{Garcia65}
{Garcia}, J.~D. \& {Mack}, J.~E. 1965, Journal of the Optical Society of
  America (1917-1983), 55, 654

\bibitem[{{Garmire} {et~al.}(2003){Garmire}, {Bautz}, {Ford}, {Nousek}, \&
  {Ricker}}]{Garmire03}
{Garmire}, G.~P., {Bautz}, M.~W., {Ford}, P.~G., {Nousek}, J.~A., \& {Ricker},
  George~R., J. 2003, in Society of Photo-Optical Instrumentation Engineers
  (SPIE) Conference Series, Vol. 4851, X-Ray and Gamma-Ray Telescopes and
  Instruments for Astronomy., ed. J.~E. {Truemper} \& H.~D. {Tananbaum}, 28--44

\bibitem[{{Gies} \& {Bolton}(1982)}]{Gies82}
{Gies}, D.~R. \& {Bolton}, C.~T. 1982, \apj, 260, 240

\bibitem[{{Gies} \& {Bolton}(1986{\natexlab{a}})}]{Gies86b}
{Gies}, D.~R. \& {Bolton}, C.~T. 1986{\natexlab{a}}, \apj, 304, 389

\bibitem[{{Gies} \& {Bolton}(1986{\natexlab{b}})}]{Gies86a}
{Gies}, D.~R. \& {Bolton}, C.~T. 1986{\natexlab{b}}, \apj, 304, 371

\bibitem[{{Gies} {et~al.}(2003){Gies}, {Bolton}, {Thomson}, {Huang}, {McSwain},
  {Riddle}, {Wang}, {Wiita}, {Wingert}, {Cs{\'a}k}, \& {Kiss}}]{Gies03}
{Gies}, D.~R., {Bolton}, C.~T., {Thomson}, J.~R., {et~al.} 2003, \apj, 583, 424

\bibitem[{{Grinberg} {et~al.}(2013){Grinberg}, {Hell}, {Pottschmidt},
  {B{\"o}ck}, {Nowak}, {Rodriguez}, {Bodaghee}, {Cadolle Bel}, {Case}, {Hanke},
  {K{\"u}hnel}, {Markoff}, {Pooley}, {Rothschild}, {Tomsick}, {Wilson-Hodge},
  \& {Wilms}}]{Grinberg13}
{Grinberg}, V., {Hell}, N., {Pottschmidt}, K., {et~al.} 2013, \aap, 554, A88

\bibitem[{{Grinberg} {et~al.}(2015){Grinberg}, {Leutenegger}, {Hell},
  {Pottschmidt}, {B{\"o}ck}, {Garc{\'\i}a}, {Hanke}, {Nowak}, {Sundqvist},
  {Townsend}, \& {Wilms}}]{Grinberg15}
{Grinberg}, V., {Leutenegger}, M.~A., {Hell}, N., {et~al.} 2015, \aap, 576,
  A117

\bibitem[{{Grinberg} {et~al.}(2020){Grinberg}, {Nowak}, \&
  {Hell}}]{Grinberg_2020a}
{Grinberg}, V., {Nowak}, M.~A., \& {Hell}, N. 2020, \aap, 643, A109

\bibitem[{{Hamann} {et~al.}(2008){Hamann}, {Feldmeier}, \&
  {Oskinova}}]{Hamann08}
{Hamann}, W.-R., {Feldmeier}, A., \& {Oskinova}, L.~M. 2008, in Clumping in
  Hot-Star Winds

\bibitem[{{Hanke} {et~al.}(2009){Hanke}, {Wilms}, {Nowak}, {Pottschmidt},
  {Schulz}, \& {Lee}}]{Hanke09}
{Hanke}, M., {Wilms}, J., {Nowak}, M.~A., {et~al.} 2009, \apj, 690, 330

\bibitem[{{H{\"a}rer} {et~al.}(2021){H{\"a}rer}, {Parker}, {Joyce}, {Igo},
  {Alston}, {F{\"u}rst}, {Lobban}, {Matzeu}, \& {Reeves}}]{Haerer21_PDS}
{H{\"a}rer}, L., {Parker}, M.~L., {Joyce}, A., {et~al.} 2021, \mnras, 500, 4506

\bibitem[{{Hell} {et~al.}(2016){Hell}, {Brown}, {Wilms}, {Grinberg},
  {Clementson}, {Liedahl}, {Porter}, {Kelley}, {Kilbourne}, \&
  {Beiersdorfer}}]{Hell16}
{Hell}, N., {Brown}, G.~V., {Wilms}, J., {et~al.} 2016, \apj, 830, 26

\bibitem[{{Herrero} {et~al.}(1995){Herrero}, {Kudritzki}, {Gabler}, {Vilchez},
  \& {Gabler}}]{Herrero95}
{Herrero}, A., {Kudritzki}, R.~P., {Gabler}, R., {Vilchez}, J.~M., \& {Gabler},
  A. 1995, \aap, 297, 556

\bibitem[{{Hirsch} {et~al.}(2019){Hirsch}, {Hell}, {Grinberg}, {Ballhausen},
  {Nowak}, {Pottschmidt}, {Schulz}, {Dauser}, {Hanke}, {Kallman}, {Brown}, \&
  {Wilms}}]{Hirsch19}
{Hirsch}, M., {Hell}, N., {Grinberg}, V., {et~al.} 2019, \aap, 626, A64

\bibitem[{{Ibragimov} {et~al.}(2005){Ibragimov}, {Poutanen}, {Gilfanov},
  {Zdziarski}, \& {Shrader}}]{Ibragimov_2005a}
{Ibragimov}, A., {Poutanen}, J., {Gilfanov}, M., {Zdziarski}, A.~A., \&
  {Shrader}, C.~R. 2005, \mnras, 362, 1435

\bibitem[{{Igo} {et~al.}(2020){Igo}, {Parker}, {Matzeu}, {Alston}, {Alvarez
  Crespo}, {F{\"u}rst}, {Buisson}, {Lobban}, {Joyce}, {Mallick}, {Schartel}, \&
  {Santos-Lle{\'o}}}]{Igo20}
{Igo}, Z., {Parker}, M.~L., {Matzeu}, G.~A., {et~al.} 2020, \mnras, 493, 1088

\bibitem[{{in't Zand}(2005)}]{in'tZand05}
{in't Zand}, J.~J.~M. 2005, \aap, 441, L1

\bibitem[{{Johnson} \& {Soff}(1985)}]{Johnson85}
{Johnson}, W.~R. \& {Soff}, G. 1985, Atomic Data and Nuclear Data Tables, 33,
  405

\bibitem[{{Kallman} \& {Bautista}(2001)}]{Kallman01}
{Kallman}, T. \& {Bautista}, M. 2001, \apjs, 133, 221

\bibitem[{{Krawczynski} {et~al.}(2022){Krawczynski}, {Muleri}, {Dov{\v{c}}iak},
  {Veledina}, {Rodriguez Cavero}, {Svoboda}, {Ingram}, {Matt}, {Garcia},
  {Loktev}, {Negro}, {Poutanen}, {Kitaguchi}, {Podgorn{\'y}}, {Rankin},
  {Zhang}, {Berdyugin}, {Berdyugina}, {Bianchi}, {Blinov}, {Capitanio}, {Di
  Lalla}, {Draghis}, {Fabiani}, {Kagitani}, {Kravtsov}, {Kiehlmann},
  {Latronico}, {Lutovinov}, {Mandarakas}, {Marin}, {Marinucci}, {Miller},
  {Mizuno}, {Molkov}, {Omodei}, {Petrucci}, {Ratheesh}, {Sakanoi}, {Semena},
  {Skalidis}, {Soffitta}, {Tennant}, {Thalhammer}, {Tombesi}, {Weisskopf},
  {Wilms}, {Zhang}, {Agudo}, {Antonelli}, {Bachetti}, {Baldini}, {Baumgartner},
  {Bellazzini}, {Bongiorno}, {Bonino}, {Brez}, {Bucciantini}, {Castellano},
  {Cavazzuti}, {Ciprini}, {Costa}, {De Rosa}, {Del Monte}, {Di Gesu}, {Di
  Marco}, {Donnarumma}, {Doroshenko}, {Ehlert}, {Enoto}, {Evangelista},
  {Ferrazzoli}, {Gunji}, {Hayashida}, {Heyl}, {Iwakiri}, {Jorstad}, {Karas},
  {Kolodziejczak}, {La Monaca}, {Liodakis}, {Maldera}, {Manfreda}, {Marscher},
  {Marshall}, {Mitsuishi}, {Ng}, {O{\textquoteright}Dell}, {Oppedisano},
  {Papitto}, {Pavlov}, {Peirson}, {Perri}, {Pesce-Rollins}, {Pilia},
  {Possenti}, {Puccetti}, {Ramsey}, {Romani}, {Sgr{\`o}}, {Slane}, {Spandre},
  {Tamagawa}, {Tavecchio}, {Taverna}, {Tawara}, {Thomas}, {Trois}, {Tsygankov},
  {Turolla}, {Vink}, {Wu}, {Xie}, \& {Zane}}]{Krawczynski22}
{Krawczynski}, H., {Muleri}, F., {Dov{\v{c}}iak}, M., {et~al.} 2022, Science,
  378, 650

\bibitem[{{Krti{\v{c}}ka} {et~al.}(2018){Krti{\v{c}}ka}, {Kub{\'a}t}, \&
  {Krti{\v{c}}kov{\'a}}}]{krticka18}
{Krti{\v{c}}ka}, J., {Kub{\'a}t}, J., \& {Krti{\v{c}}kov{\'a}}, I. 2018, \aap,
  620, A150

\bibitem[{{Liao} {et~al.}(2013){Liao}, {Zhang}, \& {Yao}}]{Liao13}
{Liao}, J.-Y., {Zhang}, S.-N., \& {Yao}, Y. 2013, \apj, 774, 116

\bibitem[{{Lucy} \& {Solomon}(1970)}]{Lucy70}
{Lucy}, L.~B. \& {Solomon}, P.~M. 1970, \apj, 159, 879

\bibitem[{{Mendoza} {et~al.}(2021){Mendoza}, {Bautista}, {Deprince},
  {Garc{\'\i}a}, {Gatuzz}, {Gorczyca}, {Kallman}, {Palmeri}, {Quinet}, \&
  {Witthoeft}}]{Mendoza21}
{Mendoza}, C., {Bautista}, M.~A., {Deprince}, J., {et~al.} 2021, Atoms, 9, 12

\bibitem[{{Miller-Jones} {et~al.}(2021){Miller-Jones}, {Bahramian}, {Orosz},
  {Mandel}, {Gou}, {Maccarone}, {Neijssel}, {Zhao}, {Zi{\'o}{\l}kowski},
  {Reid}, {Uttley}, {Zheng}, {Byun}, {Dodson}, {Grinberg}, {Jung}, {Kim},
  {Marcote}, {Markoff}, {Rioja}, {Rushton}, {Russell}, {Sivakoff}, {Tetarenko},
  {Tudose}, \& {Wilms}}]{Miller-Jones21}
{Miller-Jones}, J. C.~A., {Bahramian}, A., {Orosz}, J.~A., {et~al.} 2021,
  Science, 371, 1046

\bibitem[{{Mi{\v{s}}kovi{\v{c}}ov{\'a}}
  {et~al.}(2016){Mi{\v{s}}kovi{\v{c}}ov{\'a}}, {Hell}, {Hanke}, {Nowak},
  {Pottschmidt}, {Schulz}, {Grinberg}, {Duro}, {Madej}, {Lohfink}, {Rodriguez},
  {Cadolle Bel}, {Bodaghee}, {Tomsick}, {Lee}, {Brown}, \&
  {Wilms}}]{Miskovicova16}
{Mi{\v{s}}kovi{\v{c}}ov{\'a}}, I., {Hell}, N., {Hanke}, M., {et~al.} 2016,
  \aap, 590, A114

\bibitem[{{Mizumoto} \& {Ebisawa}(2017)}]{Mizumoto17}
{Mizumoto}, M. \& {Ebisawa}, K. 2017, \mnras, 466, 3259

\bibitem[{{Nandra} {et~al.}(2013){Nandra}, {Barret}, {Barcons}, {Fabian}, {den
  Herder}, {Piro}, {Watson}, {Adami}, {Aird}, {Afonso}, {Alexander},
  {Argiroffi}, {Amati}, {Arnaud}, {Atteia}, {Audard}, {Badenes}, {Ballet},
  {Ballo}, {Bamba}, {Bhardwaj}, {Stefano Battistelli}, {Becker}, {De Becker},
  {Behar}, {Bianchi}, {Biffi}, {B{\^\i}rzan}, {Bocchino}, {Bogdanov}, {Boirin},
  {Boller}, {Borgani}, {Borm}, {Bouch{\'e}}, {Bourdin}, {Bower}, {Braito},
  {Branchini}, {Branduardi-Raymont}, {Bregman}, {Brenneman}, {Brightman},
  {Br{\"u}ggen}, {Buchner}, {Bulbul}, {Brusa}, {Bursa}, {Caccianiga},
  {Cackett}, {Campana}, {Cappelluti}, {Cappi}, {Carrera}, {Ceballos},
  {Christensen}, {Chu}, {Churazov}, {Clerc}, {Corbel}, {Corral}, {Comastri},
  {Costantini}, {Croston}, {Dadina}, {D'Ai}, {Decourchelle}, {Della Ceca},
  {Dennerl}, {Dolag}, {Done}, {Dovciak}, {Drake}, {Eckert}, {Edge}, {Ettori},
  {Ezoe}, {Feigelson}, {Fender}, {Feruglio}, {Finoguenov}, {Fiore}, {Galeazzi},
  {Gallagher}, {Gandhi}, {Gaspari}, {Gastaldello}, {Georgakakis},
  {Georgantopoulos}, {Gilfanov}, {Gitti}, {Gladstone}, {Goosmann}, {Gosset},
  {Grosso}, {Guedel}, {Guerrero}, {Haberl}, {Hardcastle}, {Heinz}, {Alonso
  Herrero}, {Herv{\'e}}, {Holmstrom}, {Iwasawa}, {Jonker}, {Kaastra}, {Kara},
  {Karas}, {Kastner}, {King}, {Kosenko}, {Koutroumpa}, {Kraft}, {Kreykenbohm},
  {Lallement}, {Lanzuisi}, {Lee}, {Lemoine-Goumard}, {Lobban}, {Lodato},
  {Lovisari}, {Lotti}, {McCharthy}, {McNamara}, {Maggio}, {Maiolino}, {De
  Marco}, {de Martino}, {Mateos}, {Matt}, {Maughan}, {Mazzotta}, {Mendez},
  {Merloni}, {Micela}, {Miceli}, {Mignani}, {Miller}, {Miniutti}, {Molendi},
  {Montez}, {Moretti}, {Motch}, {Naz{\'e}}, {Nevalainen}, {Nicastro}, {Nulsen},
  {Ohashi}, {O'Brien}, {Osborne}, {Oskinova}, {Pacaud}, {Paerels}, {Page},
  {Papadakis}, {Pareschi}, {Petre}, {Petrucci}, {Piconcelli}, {Pillitteri},
  {Pinto}, {de Plaa}, {Pointecouteau}, {Ponman}, {Ponti}, {Porquet}, {Pounds},
  {Pratt}, {Predehl}, {Proga}, {Psaltis}, {Rafferty}, {Ramos-Ceja}, {Ranalli},
  {Rasia}, {Rau}, {Rauw}, {Rea}, {Read}, {Reeves}, {Reiprich}, {Renaud},
  {Reynolds}, {Risaliti}, {Rodriguez}, {Rodriguez Hidalgo}, {Roncarelli},
  {Rosario}, {Rossetti}, {Rozanska}, {Rovilos}, {Salvaterra}, {Salvato}, {Di
  Salvo}, {Sanders}, {Sanz-Forcada}, {Schawinski}, {Schaye}, {Schwope},
  {Sciortino}, {Severgnini}, {Shankar}, {Sijacki}, {Sim}, {Schmid}, {Smith},
  {Steiner}, {Stelzer}, {Stewart}, {Strohmayer}, {Str{\"u}der}, {Sun}, {Takei},
  {Tatischeff}, {Tiengo}, {Tombesi}, {Trinchieri}, {Tsuru}, {Ud-Doula},
  {Ursino}, {Valencic}, {Vanzella}, {Vaughan}, {Vignali}, {Vink}, {Vito},
  {Volonteri}, {Wang}, {Webb}, {Willingale}, {Wilms}, {Wise}, {Worrall},
  {Young}, {Zampieri}, {In't Zand}, {Zane}, {Zezas}, {Zhang}, \&
  {Zhuravleva}}]{Nandra13}
{Nandra}, K., {Barret}, D., {Barcons}, X., {et~al.} 2013, arXiv e-prints,
  arXiv:1306.2307

\bibitem[{{Ness} {et~al.}(2012){Ness}, {Schaefer}, {Dobrotka}, {Sadowski},
  {Drake}, {Barnard}, {Talavera}, {Gonzalez-Riestra}, {Page}, {Hernanz},
  {Sala}, \& {Starrfield}}]{Ness12}
{Ness}, J.~U., {Schaefer}, B.~E., {Dobrotka}, A., {et~al.} 2012, \apj, 745, 43

\bibitem[{{Nowak} {et~al.}(2011){Nowak}, {Hanke}, {Trowbridge}, {Markoff},
  {Wilms}, {Pottschmidt}, {Coppi}, {Maitra}, {Davis}, \&
  {Tramper}}]{Nowak_2011a}
{Nowak}, M.~A., {Hanke}, M., {Trowbridge}, S.~N., {et~al.} 2011, \apj, 728, 13

\bibitem[{{Orosz} {et~al.}(2011){Orosz}, {McClintock}, {Aufdenberg},
  {Remillard}, {Reid}, {Narayan}, \& {Gou}}]{Orosz11}
{Orosz}, J.~A., {McClintock}, J.~E., {Aufdenberg}, J.~P., {et~al.} 2011, \apj,
  742, 84

\bibitem[{{Oskinova} {et~al.}(2012){Oskinova}, {Feldmeier}, \&
  {Kretschmar}}]{Oskinova12}
{Oskinova}, L.~M., {Feldmeier}, A., \& {Kretschmar}, P. 2012, \mnras, 421, 2820

\bibitem[{{Owocki} {et~al.}(1988){Owocki}, {Castor}, \& {Rybicki}}]{Owocki88}
{Owocki}, S.~P., {Castor}, J.~I., \& {Rybicki}, G.~B. 1988, \apj, 335, 914

\bibitem[{{Owocki} \& {Rybicki}(1984)}]{Owocki84}
{Owocki}, S.~P. \& {Rybicki}, G.~B. 1984, \apj, 284, 337

\bibitem[{{Parker} {et~al.}(2017{\natexlab{a}}){Parker}, {Alston}, {Buisson},
  {Fabian}, {Jiang}, {Kara}, {Lohfink}, {Pinto}, \& {Reynolds}}]{Parker17_iras}
{Parker}, M.~L., {Alston}, W.~N., {Buisson}, D.~J.~K., {et~al.}
  2017{\natexlab{a}}, \mnras, 469, 1553

\bibitem[{{Parker} {et~al.}(2020){Parker}, {Alston}, {Igo}, \&
  {Fabian}}]{Parker20}
{Parker}, M.~L., {Alston}, W.~N., {Igo}, Z., \& {Fabian}, A.~C. 2020, \mnras,
  492, 1363

\bibitem[{{Parker} {et~al.}(2017{\natexlab{b}}){Parker}, {Pinto}, {Fabian},
  {Lohfink}, {Buisson}, {Alston}, {Kara}, {Cackett}, {Chiang}, {Dauser}, {De
  Marco}, {Gallo}, {Garcia}, {Harrison}, {King}, {Middleton}, {Miller},
  {Miniutti}, {Reynolds}, {Uttley}, {Vasudevan}, {Walton}, {Wilkins}, \&
  {Zoghbi}}]{Parker17_nature}
{Parker}, M.~L., {Pinto}, C., {Fabian}, A.~C., {et~al.} 2017{\natexlab{b}},
  \nat, 543, 83

\bibitem[{{Parker} {et~al.}(2018){Parker}, {Reeves}, {Matzeu}, {Buisson}, \&
  {Fabian}}]{Parker18_PCA}
{Parker}, M.~L., {Reeves}, J.~N., {Matzeu}, G.~A., {Buisson}, D.~J.~K., \&
  {Fabian}, A.~C. 2018, \mnras, 474, 108

\bibitem[{{Parker} {et~al.}(2015){Parker}, {Tomsick}, {Miller}, {Yamaoka},
  {Lohfink}, {Nowak}, {Fabian}, {Alston}, {Boggs}, {Christensen}, {Craig},
  {F{\"u}rst}, {Gandhi}, {Grefenstette}, {Grinberg}, {Hailey}, {Harrison},
  {Kara}, {King}, {Stern}, {Walton}, {Wilms}, \& {Zhang}}]{Parker_2015a}
{Parker}, M.~L., {Tomsick}, J.~A., {Miller}, J.~M., {et~al.} 2015, \apj, 808, 9

\bibitem[{{Perucho} \& {Bosch-Ramon}(2012)}]{Perucho12}
{Perucho}, M. \& {Bosch-Ramon}, V. 2012, \aap, 539, A57

\bibitem[{{Pinto} {et~al.}(2018){Pinto}, {Alston}, {Parker}, {Fabian}, {Gallo},
  {Buisson}, {Walton}, {Kara}, {Jiang}, {Lohfink}, \& {Reynolds}}]{Pinto18}
{Pinto}, C., {Alston}, W., {Parker}, M.~L., {et~al.} 2018, \mnras, 476, 1021

\bibitem[{{Puls} {et~al.}(2015){Puls}, {Sundqvist}, \& {Markova}}]{Puls15}
{Puls}, J., {Sundqvist}, J.~O., \& {Markova}, N. 2015, in New Windows on
  Massive Stars, ed. G.~{Meynet}, C.~{Georgy}, J.~{Groh}, \& P.~{Stee}, Vol.
  307, 25--36

\bibitem[{{Puls} {et~al.}(2008){Puls}, {Vink}, \& {Najarro}}]{Puls08}
{Puls}, J., {Vink}, J.~S., \& {Najarro}, F. 2008, \aapr, 16, 209

\bibitem[{{Sander} {et~al.}(2018){Sander}, {F{\"u}rst}, {Kretschmar},
  {Oskinova}, {Todt}, {Hainich}, {Shenar}, \& {Hamann}}]{Sander18}
{Sander}, A.~A.~C., {F{\"u}rst}, F., {Kretschmar}, P., {et~al.} 2018, \aap,
  610, A60

\bibitem[{{Sander} {et~al.}(2017){Sander}, {Hamann}, {Todt}, {Hainich}, \&
  {Shenar}}]{Sander17}
{Sander}, A.~A.~C., {Hamann}, W.~R., {Todt}, H., {Hainich}, R., \& {Shenar}, T.
  2017, \aap, 603, A86

\bibitem[{{Sundqvist} {et~al.}(2019){Sundqvist}, {Bj{\"o}rklund}, {Puls}, \&
  {Najarro}}]{Sundqvist19}
{Sundqvist}, J.~O., {Bj{\"o}rklund}, R., {Puls}, J., \& {Najarro}, F. 2019,
  \aap, 632, A126

\bibitem[{{Sundqvist} {et~al.}(2012){Sundqvist}, {Owocki}, \&
  {Puls}}]{Sundqvist12}
{Sundqvist}, J.~O., {Owocki}, S.~P., \& {Puls}, J. 2012, in Astronomical
  Society of the Pacific Conference Series, Vol. 465, Proceedings of a
  Scientific Meeting in Honor of Anthony F. J. Moffat, ed. L.~{Drissen},
  C.~{Robert}, N.~{St-Louis}, \& A.~F.~J. {Moffat}, 119

\bibitem[{{Sundqvist} {et~al.}(2018){Sundqvist}, {Owocki}, \&
  {Puls}}]{Sundqvist18a}
{Sundqvist}, J.~O., {Owocki}, S.~P., \& {Puls}, J. 2018, \aap, 611, A17

\bibitem[{{Sundqvist} \& {Puls}(2018)}]{Sundqvist18b}
{Sundqvist}, J.~O. \& {Puls}, J. 2018, \aap, 619, A59

\bibitem[{{Tarter} {et~al.}(1969){Tarter}, {Tucker}, \& {Salpeter}}]{Tarter69}
{Tarter}, C.~B., {Tucker}, W.~H., \& {Salpeter}, E.~E. 1969, \apj, 156, 943

\bibitem[{{Tashiro} {et~al.}(2018){Tashiro}, {Maejima}, {Toda}, {Kelley},
  {Reichenthal}, {Lobell}, {Petre}, {Guainazzi}, {Costantini}, {Edison},
  {Fujimoto}, {Grim}, {Hayashida}, {den Herder}, {Ishisaki}, {Paltani},
  {Matsushita}, {Mori}, {Sneiderman}, {Takei}, {Terada}, {Tomida}, {Akamatsu},
  {Angelini}, {Arai}, {Awaki}, {Babyk}, {Bamba}, {Barfknecht}, {Barnstable},
  {Bialas}, {Blagojevic}, {Bonafede}, {Brambora}, {Brenneman}, {Brown},
  {Brown}, {Burns}, {Canavan}, {Carnahan}, {Chiao}, {Comber}, {Corrales}, {de
  Vries}, {Dercksen}, {Diaz-Trigo}, {Dillard}, {DiPirro}, {Done}, {Dotani},
  {Ebisawa}, {Eckart}, {Enoto}, {Ezoe}, {Ferrigno}, {Fukazawa}, {Fujita},
  {Furuzawa}, {Gallo}, {Graham}, {Gu}, {Hagino}, {Hamaguchi}, {Hatsukade},
  {Hawes}, {Hayashi}, {Hegarty}, {Hell}, {Hiraga}, {Hodges-Kluck}, {Holland},
  {Hornschemeier}, {Hoshino}, {Ichinohe}, {Iizuka}, {Ishibashi}, {Ishida},
  {Ishikawa}, {Ishimura}, {James}, {Kallman}, {Kara}, {Katsuda}, {Kenyon},
  {Kilbourne}, {Kimball}, {Kitaguti}, {Kitamoto}, {Kobayashi}, {Kohmura},
  {Koyama}, {Kubota}, {Leutenegger}, {Lockard}, {Loewenstein}, {Maeda},
  {Marbley}, {Markevitch}, {Matsumoto}, {Matsuzaki}, {McCammon}, {McNamara},
  {Miko}, {Miller}, {Miller}, {Minesugi}, {Mitsuishi}, {Mizuno}, {Mori},
  {Mukai}, {Murakami}, {Mushotzky}, {Nakajima}, {Nakamura}, {Nakashima},
  {Nakazawa}, {Natsukari}, {Nigo}, {Nishioka}, {Nobukawa}, {Nobukawa}, {Noda},
  {Odaka}, {Ogawa}, {Ohashi}, {Ohno}, {Ohta}, {Okajima}, {Okamoto}, {Onizuka},
  {Ota}, {Ozaki}, {Plucinsky}, {Porter}, {Pottschmidt}, {Sato}, {Sato},
  {Sawada}, {Seta}, {Shelton}, {Shibano}, {Shida}, {Shidatsu}, {Shirron},
  {Simionescu}, {Smith}, {Someya}, {Soong}, {Suagawara}, {Szymkowiak},
  {Takahashi}, {Tamagawa}, {Tamura}, {Tanaka}, {Terashima}, {Tsuboi},
  {Tsujimoto}, {Tsunemi}, {Tsuru}, {Uchida}, {Uchiyama}, {Ueda}, {Uno},
  {Walsh}, {Watanabe}, {Williams}, {Wolfs}, {Wright}, {Yamada}, {Yamaguchi},
  {Yamaoka}, {Yamasaki}, {Yamauchi}, {Yamauchi}, {Yanagase}, {Yaqoob},
  {Yasuda}, {Yoshioka}, {Zabala}, \& {Irina}}]{Tashiro18}
{Tashiro}, M., {Maejima}, H., {Toda}, K., {et~al.} 2018, in Society of
  Photo-Optical Instrumentation Engineers (SPIE) Conference Series, Vol. 10699,
  Space Telescopes and Instrumentation 2018: Ultraviolet to Gamma Ray, ed.
  J.-W.~A. {den Herder}, S.~{Nikzad}, \& K.~{Nakazawa}, 1069922

\bibitem[{{Tombesi} {et~al.}(2010){Tombesi}, {Cappi}, {Yaqoob}, {Reeves},
  {Braito}, \& {Palumbo}}]{Tombesi10}
{Tombesi}, F., {Cappi}, M., {Yaqoob}, T., {et~al.} 2010, in Astronomical
  Society of the Pacific Conference Series, Vol. 427, Accretion and Ejection in
  AGN: a Global View, ed. L.~{Maraschi}, G.~{Ghisellini}, R.~{Della Ceca}, \&
  F.~{Tavecchio}, 120

\bibitem[{{Tomsick} {et~al.}(2014){Tomsick}, {Nowak}, {Parker}, {Miller},
  {Fabian}, {Harrison}, {Bachetti}, {Barret}, {Boggs}, {Christensen}, {Craig},
  {Forster}, {F{\"u}rst}, {Grefenstette}, {Hailey}, {King}, {Madsen},
  {Natalucci}, {Pottschmidt}, {Ross}, {Stern}, {Walton}, {Wilms}, \&
  {Zhang}}]{Tomsick_2014a}
{Tomsick}, J.~A., {Nowak}, M.~A., {Parker}, M., {et~al.} 2014, \apj, 780, 78

\bibitem[{{Tomsick} {et~al.}(2018){Tomsick}, {Parker}, {Garc{\'\i}a},
  {Yamaoka}, {Barret}, {Chiu}, {Clavel}, {Fabian}, {F{\"u}rst}, {Gandhi},
  {Grinberg}, {Miller}, {Pottschmidt}, \& {Walton}}]{Tomsick_2018a}
{Tomsick}, J.~A., {Parker}, M.~L., {Garc{\'\i}a}, J.~A., {et~al.} 2018, \apj,
  855, 3

\bibitem[{{Vaughan} {et~al.}(2003){Vaughan}, {Edelson}, {Warwick}, \&
  {Uttley}}]{Vaughan03}
{Vaughan}, S., {Edelson}, R., {Warwick}, R.~S., \& {Uttley}, P. 2003, \mnras,
  345, 1271

\bibitem[{{Verner} {et~al.}(1996){Verner}, {Verner}, \& {Ferland}}]{Verner96}
{Verner}, D.~A., {Verner}, E.~M., \& {Ferland}, G.~J. 1996, Atomic Data and
  Nuclear Data Tables, 64, 1

\bibitem[{{Walborn}(1973)}]{Walborn73}
{Walborn}, N.~R. 1973, \apjl, 179, L123

\bibitem[{{Walton} {et~al.}(2016){Walton}, {Tomsick}, {Madsen}, {Grinberg},
  {Barret}, {Boggs}, {Christensen}, {Clavel}, {Craig}, {Fabian}, {Fuerst},
  {Hailey}, {Harrison}, {Miller}, {Parker}, {Rahoui}, {Stern}, {Tao}, {Wilms},
  \& {Zhang}}]{Walton_2016a}
{Walton}, D.~J., {Tomsick}, J.~A., {Madsen}, K.~K., {et~al.} 2016, \apj, 826,
  87

\bibitem[{{Webster} \& {Murdin}(1972)}]{Webster72}
{Webster}, B.~L. \& {Murdin}, P. 1972, \nat, 235, 37

\bibitem[{{Wilms} {et~al.}(2006){Wilms}, {Nowak}, {Pottschmidt}, {Pooley}, \&
  {Fritz}}]{Wilms_2006a}
{Wilms}, J., {Nowak}, M.~A., {Pottschmidt}, K., {Pooley}, G.~G., \& {Fritz}, S.
  2006, \aap, 447, 245

\end{thebibliography}
\bibliographystyle{aa}



\begin{appendix}
\section{Filtering CCD gaps and dead pixel rows}
\label{sec:gapfiltering}

In Sect.~\ref{sec:obs_meth}, we briefly discussed that the dither in \chandra\ introduces a systematic increase in \fvsq\ around CCD gaps and rows of dead pixels. In this appendix, we explain in detail how affected wavelengths bins are identified.

\begin{figure*}
    \centering
    \includegraphics[width = 0.95\textwidth]{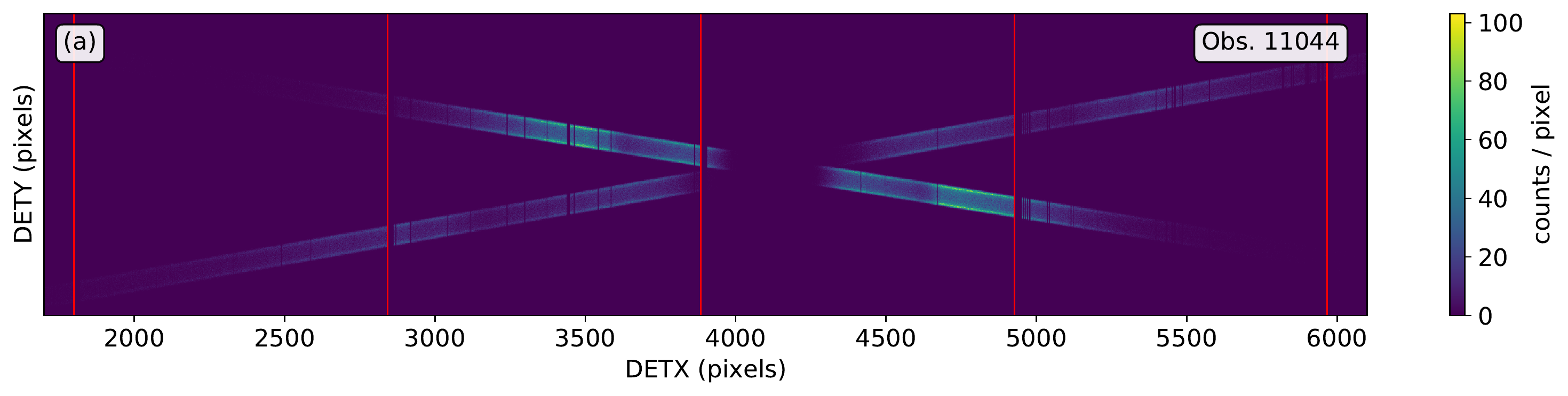}
    \includegraphics[width = 0.95\textwidth]{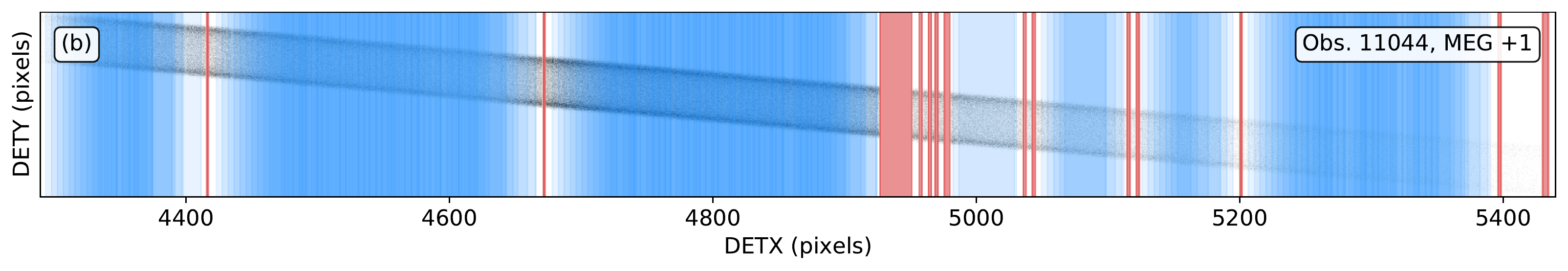}
    \includegraphics[width = 0.95\textwidth]{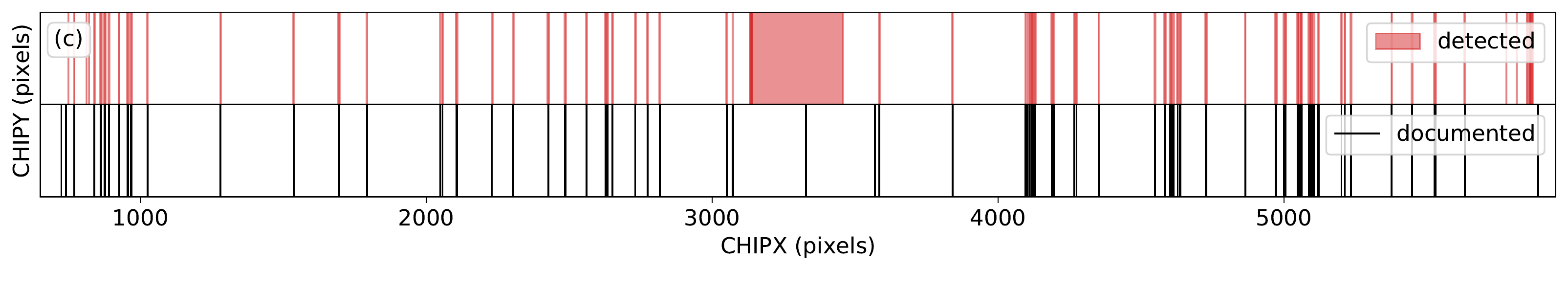}
    \includegraphics[width = 0.95\textwidth]{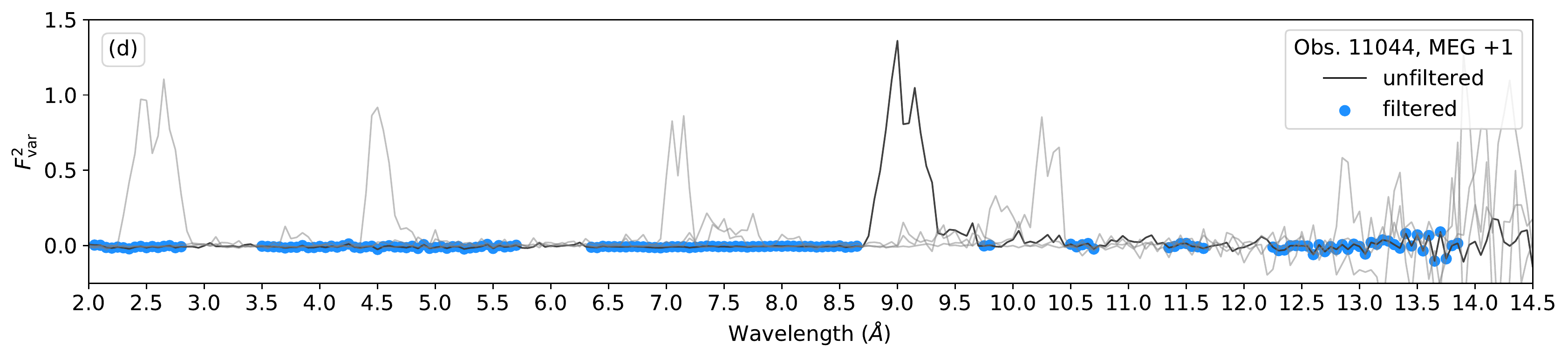}
    \includegraphics[width = 0.95\textwidth]{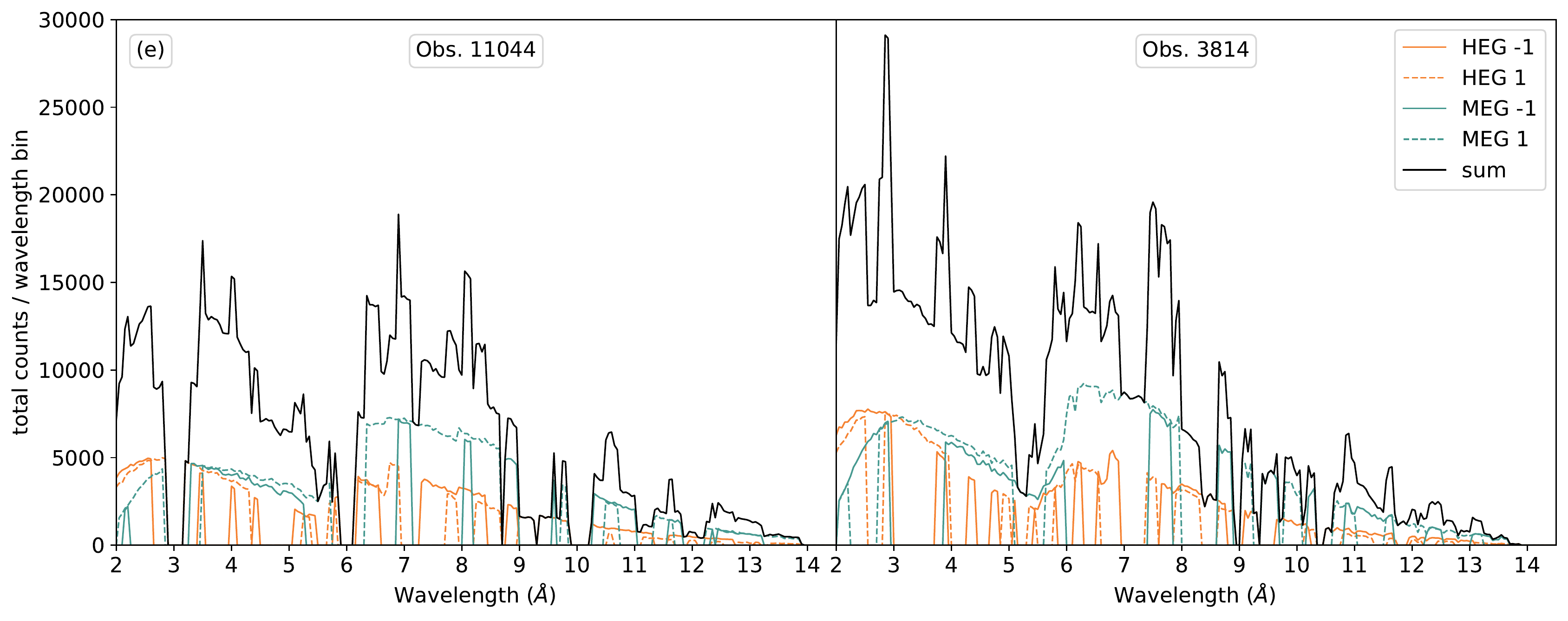}
    \caption{CCD gaps and dead pixel rows in \chandra's ACIS-S detector are a major source of systematic variance. This figure illustrates our filtering procedure. (a) The distribution of counts on the ACIS-S CCD array, against which CCD gaps (marked in red) and dead pixel rows clearly stand out. Panel (b) shows the filtered bins (blue) and gaps (red) as an overlay on a cut-out of the top panel (first order MEG spectrum). Only bins with a minimum distance of two pixels to the detected gaps pass the filtering. Panel (c) compares the positions of detected gaps to those of dead pixel rows documented in the ACIS \texttt{badpix} calibration file. The gap at ${\sim}3300\,$pixels is not excluded because of a detector defect, but because it contains zeroth order and $\lambda < 2$\AA\ events. In panel (d), the resulting \fvsq\ is shown, with the filtered bins highlighted in blue and MEG $-1$ and HEG $\pm1$ spectra (all light gray) for comparison. Details on the filtering procedure can be found in the text. (e) Total counts per wavelength bin in Obs.~11044 (left) and 3814 (right) after filtering, for all gratings arms (orange and green, straight and dashed lines) and their sum (black line). For clarity, data in panels (d) and (e) are displayed with lines and no errors are given.}
    \label{fig:gaps}
\end{figure*}

Figure~\ref{fig:gaps}\,(a) shows the distribution of counts on \chandra's ACIS-S detector for Obs.~11044. The ACIS-S consists of a row of six quadratic CCDs, on which photons are dispersed by the HEG and MEG, resulting in a cross-like shape. Due to this design, a small amount of photons are lost at the boundaries between CCDs, in the so-called CCD gaps \citep{Garmire03}. Rows of dead pixels have the same effect. We jointly refer to CCD gaps and dead rows as gaps in the following. To avoid the photon loss exclusively affecting specific wavelengths, the telescope pointing moves in a Lissajous pattern, periodically changing the zeroth order position on the detector, which averages the loss over a small wavelength range \citep{Garmire03}. As a secondary effect, the number of detected photons strongly varies with time in range of a gap, which strongly affects variability analyses. The diagonal positioning of the gratings increases the affected range, because it causes one gap to affect a range of wavelengths at any given moment. 

To account for the systematic variability introduced by the dither, we exclude all bins within dither range of gaps. The first steps towards identifying these bins is finding the positions of all gaps, by searching for pixel rows with zero counts along the $x$-coordinate of the detector, considering all events within 2--14.5\AA. In this range, the number of counts per pixel is high enough to conclude that pixel rows with zero counts are likely an instrumental effect and not due to low signal. The detected gaps run across both the HEG and MEG spectrum, also indicating that they are instrumental. 

A total of $\sim70$ coinciding gaps was detected in Obs.~11044 and 3814, including the CCD gaps, whose positions were correctly reproduced. Eight additional gaps were picked up be the algorithm in Obs.~11044, but not in 3814, despite being visible by eye at coinciding positions. We therefore add these gaps in by hand in Obs.~3814. The detected gaps were verified by comparing them to the positions of dead rows documented in the ACIS \texttt{badpix} calibration file. As shown in Fig.~\ref{sec:gapfiltering}~(c), all dead rows in the investigated range were found and have matching width. Note that the coloured region centred at $\sim3300\,$pixels is excluded from the analysis, because it contains zeroth order and $\lambda < 2$\AA\ events, not because of a defective chip section.

The second step is to identify all wavelength bins which are in dither range of gaps. For each gratings arm, we calculate the minimal and maximal $x$ position of all events in a given bin and check if the closest gap is at least a two pixel distance below the minimal or above the maximal value. If not, the bin is discarded. By requiring a distance of at least two pixels, we correctly reject bins that are half covered by a gap, which cannot be distinguished from bins directly adjacent to a gap. 

As a example, a result of this selection is shown in Fig.~\ref{fig:gaps}\,(b) and (d) for the first order MEG spectrum. In panel (b), the spread of all bins that passed the filtering is shown in transparent blue and the gaps are marked in red. As intended, the filtered bins do not overlap with the gaps. Panel (d) compares the filtered and unfiltered \fvsq\ spectra. CCD gaps cause huge, often double-peaked spikes. The shape results from the fact that the amount of arriving photons varies more strongly for bins around the gap edge than for bins largely covered the gap. Dead pixel rows give rise to smaller features because they are narrower. The filtering algorithm succeeds in cutting out the large CCD gap spike, as well as several smaller ones caused by the gaps indicated in the figure above. 

Figure~\ref{fig:gaps}\,(e) shows the effect of filtering on the number of counts available for the analysis in each wavelength bin. Depending on how the filtered wavelengths align across the gratings arms, the number of counts fluctuates greatly. In a small number of bins no signal is left, which results in the gaps in the data in Fig.~\ref{fig:rms}. Note that the slightly visible two hump structure is a result of \chandra's effective area \citep[compare to, e.g.,][]{Canizares05} and that the higher number of counts in Obs.~3814 is due to a higher exposure time.

\section{Estimating binning variability}
\label{sec:binvar}

The \chandra\ ACIS-S detector has a read-out time of the order of seconds \citep{Garmire03}, which not only limits its timing resolution, but also introduces systematic variability if the data are binned. Intuitively, some photons are assigned to time bin $n+1$ instead of $n$, because of the delay between photon arrival and read-out. 

We estimate the magnitude of this effect with Monte Carlo Simulations. We generate a random constant light curve by drawing photon arrival times from a constant distribution, separately for each observation, order, and grating, with their respective count numbers and exposure times. The read-out is simulated by rounding up the arrival times to the next larger time stamp present in the data, imprinting the read-out times from the observation on the simulated data. Note that the average count rates of \src\ in Obs.~11044 and 3814 are $\sim90$ and $\sim 70\,$cts$\,\mathrm{s}^{-1}$, respectively, which makes it very unlikely that this approach misses read-out cycles because no photons arrived. We confirm this assumption by checking that there are no significant outliers in the time differences between consecutive events that exceed the read-out time. For each observation, order, and grating, we repeat this procedure 100 times and average the resulting total \fvsq. Combining all gratings and orders as described in Sect.~\ref{sec:sysvar} yields the systematic binning variability, \fvsqbin. The one sigma error on \fvsqbin\ is below $2\cdot10^{-4}$, for both Obs.~3814 and 11044.

\begin{figure}
    \centering
    \includegraphics[width = \linewidth]{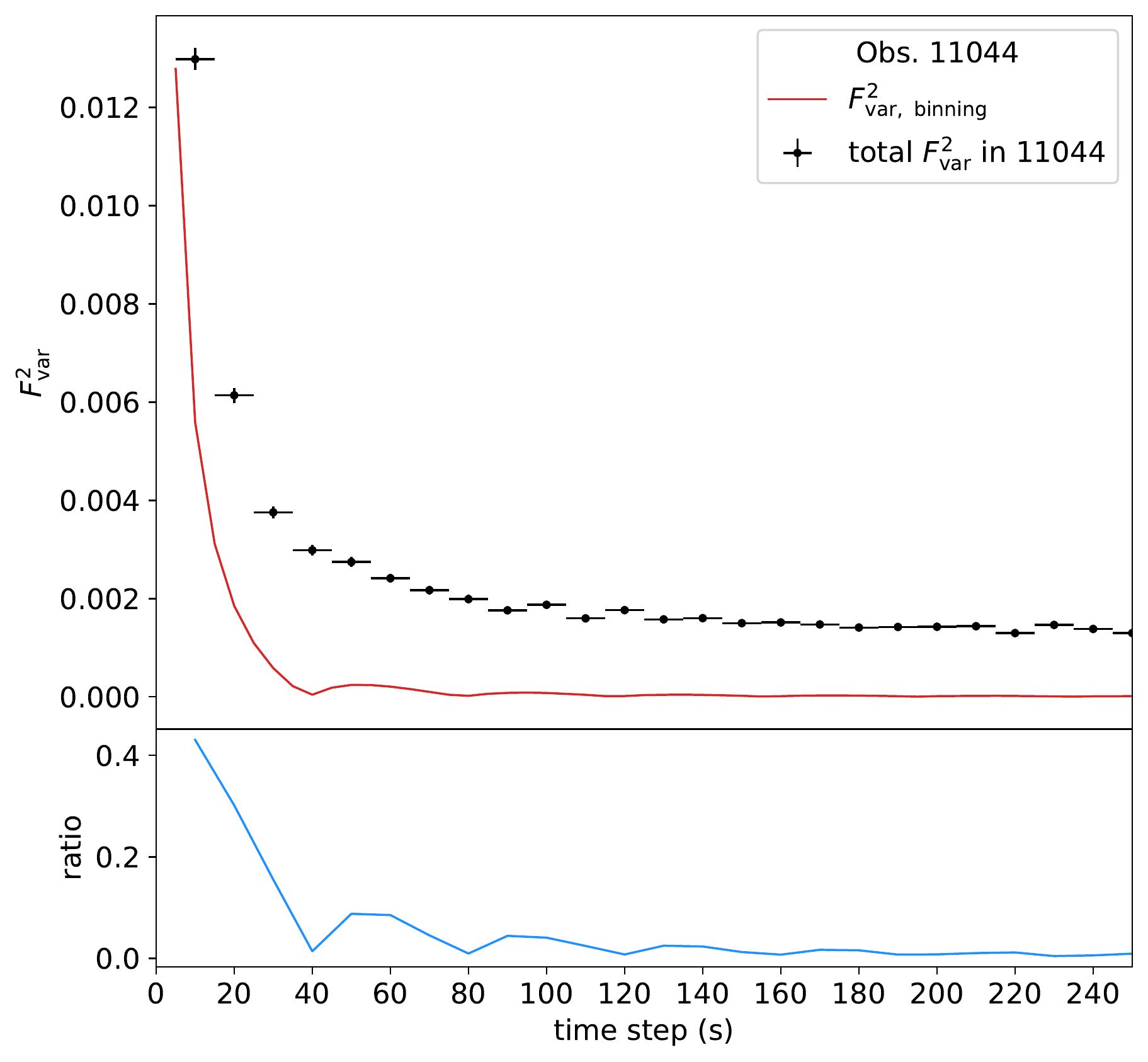}
    \caption{Systematic variability, \fvsqbin, introduced by the timing resolution of the ACIS detector (red line) compared to total \fvsq\ in the broad band ($2\mbox{--}14$\AA, black data points) and their ratio, shown in the bottom panel. As a time step of $500\,$s is used for Obs.~11044, the effect is negligible in our analysis.}
    \label{fig:binvar}
\end{figure}

An example of the simulation for Obs.~11044 is shown in Fig.~\ref{fig:binvar} as a function of binning time step. \fvsqbin\ strongly increases as the bin size approaches the read-out time, which is expected, as a smaller binning is more sensitive to a disturbance of a fixed magnitude. The periodicity in the ratio is caused by the binning time step matching an integer multiple of the duration of a read-out cycle. For a time step of 500$\,$s, which was chosen for Fig.~\ref{fig:rms}, \fvsqbin\ is below $0.9\,\%$ of the total variability over the $2\mbox{--}14$\AA\ band in Obs.~11044. For Obs.~3814, we investigate time steps as short as $50\,$s, but due to the higher variability of this observation, the fraction stays below $0.3\,\%$. In absolute terms, \fvsqbin\ always stays below $2.5\cdot10^{-5}$ for time steps above 500$\,$s in Obs.~11044 and below $2.7\cdot10^{-4}$ for time steps above 50$\,$s in Obs.~3814 at a one sigma confidence. 

We therefore conclude that \fvsqbin\ does not significantly impact our analysis, but note that the effect generally should be considered when investigating the excess variance at a timing resolution approaching the read-out time of the instrument.

\section{Detailed \fvsq\ spectrum of Obs.~11044}

\begin{table}
	\centering
	\caption{List of spike features in the \fvsq\ spectrum of Obs.~3814 with a significance ${>}2\sigma$ and associated lines, i.e., lines which have a distance of ${<}0.05$\AA\ to the features. The listed lines are compiled from the selection described in Sect.~\ref{sec:fe-select}. An alignment of lines detected in the previous analysis of the data set by \citet{Miskovicova16} was searched for, but not found. References are given below.}
	\begin{tabular}{c c l l c c}
		\hline
		\hline
		Spike [\AA] & $\sigma$ & Ion & Line & Wl. [\AA] & Ref.\\
		\hline
		5.82   &  2.2 & \ion{P}{xiv} & f [em.] & 5.836 & \citetalias{Drake88}\\
		&& \ion{P}{xiv} & i [em.] & 5.790 & \citetalias{Drake88}\\
		\hline
		8.23 & 2.0 & \ion{Fe}{xxii} & 2p$^2$ $\rightarrow$ 2p5d & 8.283 & 1\\
		&& \ion{Fe}{xxiv} & 2p $\rightarrow$ 4d & 8.232 & 1\\
		\hline
		8.58 & 2.2 & \ion{Fe}{xxi} & 2p$^2$ $\rightarrow$ 2p5d & 8.608 & 1 \\
		&&\ion{Fe}{xxiii} & 2p $\rightarrow$ 4d & 8.604 & 1 \\
		&&\ion{Fe}{xxi} & 2p$^2$ $\rightarrow$ 2p5d & 8.555 & 1 \\
		&&\ion{Fe}{xxi} & 2p$^2$ $\rightarrow$ 2p5d & 8.553 & 1 \\
		&&\ion{Fe}{xxiii} & 2p $\rightarrow$ 4d & 8.540 & 1\\
		&&\ion{Fe}{xxiii} & 2p $\rightarrow$ 4d & 8.539 & 1\\
		\hline
		11.50 & 2.1 & \ion{Ne}{ix} & 1s $\rightarrow$ 3p & 11.546 & \citetalias{Liao13} \\
		&&\ion{Fe}{xviii} & 2p$^5$ $\rightarrow$ 2p$^4$4d1 & 11.539 & 1 \\
		&&\ion{Fe}{xviii} & 2p$^5$ $\rightarrow$ 2p$^4$4d & 11.524 & 1 \\
		&&\ion{Fe}{xviii} & 2p$^5$ $\rightarrow$ 2p$^4$4d & 11.511 & 1 \\
		&&\ion{Fe}{xviii} & 2p$^5$ $\rightarrow$ 2p$^4$4d & 11.508 & 1 \\
		&&\ion{Fe}{xxii} & 2s$^2$2p $\rightarrow$ 2s2p3p & 11.480 & 1 \\
		&&\ion{Fe}{xxii} & 2s$^2$2p $\rightarrow$ 2s2p3p & 11.471 & 1 \\
		&&\ion{Fe}{xxii} & 2s$^2$2p $\rightarrow$ 2s2p3p & 11.460 & 1\\
		\hline
		\hline
		\vspace{0.5em} 
	\end{tabular}
	\label{tab:spikes_11044}
	\caption*{\textbf{References}: (1) AtomDB (\url{http://www.atomdb.org/}), \citepalias{Drake88} \citealt{Drake88}, \citepalias{Liao13} \citealt{Liao13}. \textbf{Notes}: Lines listed multiple times with different wavelengths differ in the fine structure. Emission lines are marked with [em.]. Iron lines are exclusively from AtomDB.}
\end{table}

\begin{sidewaysfigure*}
    \centering
    \includegraphics[width =0.9\textwidth]{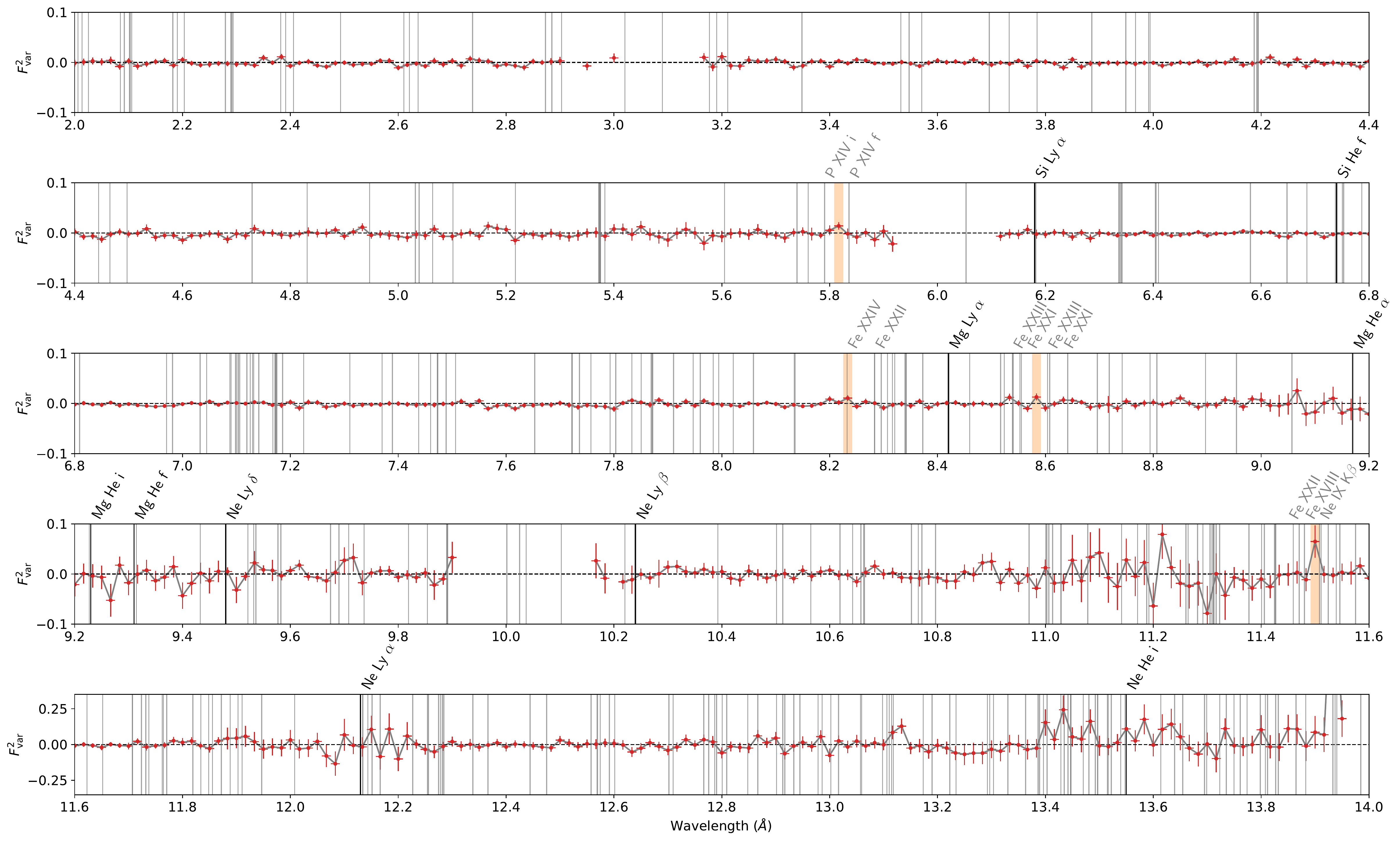}
    \caption{Detailed \fvsq\ spectrum of Obs.~11044, overplotted with the positions of line features identified in \citet{Miskovicova16} (black) and transitions from the selection described in Sect.~\ref{sec:fe-select} (grey). Spikes with a significance $\sigma > 2\sigma$ are highlighted in orange. For subsequent iron transitions belonging to the same ion the label is shown only once. A list of spikes and associated transitions can be found in Tab.~\ref{tab:spikes_11044}. In contrast to Obs.~3814, the lines indentified by \citet{Miskovicova16} are P-Cygni profiles, which makes the absence of spike features plausible (see Sect.~\ref{sec:linevar_infconj}).}
    \label{fig:11044}
\end{sidewaysfigure*}

\end{appendix}
\end{document}